\documentclass[aps,prb,twocolumn,10pt, showkeys,showpacs,amsmath,amsfonts,floatfix,superscriptaddress,nofootinbib,longbibliography]{revtex4-2}

\usepackage{url}
\renewcommand{\url}[1]{}

\usepackage{blindtext}
\usepackage{graphicx}
\usepackage{dcolumn}
\usepackage{bm}

\usepackage{hyperref}
\usepackage{physics}

\usepackage{blindtext,enumitem,dcolumn,color,xcolor,amsmath,braket,bm,changes,chemformula,cleveref,times}

\usepackage{amsmath}
\usepackage{changes}
\usepackage{multirow}
\usepackage{hhline}
\usepackage{array}

\newcommand{\br}{\bm{r}} 
\newcommand{\bk}{\bm{k}} 

\crefname{equation}{Eq.}{Eqs.}
\Crefname{equation}{Eq.}{Eqs.}
\crefname{figure}{Fig.}{Figs.}
\Crefname{figure}{Fig.}{Figs.}
\crefname{section}{Sec.}{Secs.}
\Crefname{section}{Sec.}{Secs.}
\crefname{table}{Tbl.}{Tbls.}
\Crefname{table}{Tbl.}{Tbls.}
\crefname{appendix}{App.}{Apps.}
\Crefname{appendix}{App.}{Apps.}

\begin{document}

\title{Cooper Condensation and Pair Wave Functions in Strongly Correlated Electrons}%

\author{Hannes Karlsson}
\email{hankarls@pks.mpg.de}
\affiliation{Max Planck Institute for the Physics of Complex Systems, N\"othnitzer Strasse 38, Dresden 01187, Germany}

\author{Johannes S. Hofmann}
\affiliation{Max Planck Institute for the Physics of Complex Systems, N\"othnitzer Strasse 38, Dresden 01187, Germany}

\author{Alexander Wietek}
\email{awietek@pks.mpg.de}
\affiliation{Max Planck Institute for the Physics of Complex Systems, N\"othnitzer Strasse 38, Dresden 01187, Germany}

\date{January 26, 2026}

\begin{abstract}
Identifying superconducting states of matter without prior assumptions is a central challenge in strongly correlated electron systems. We introduce a canonical framework for diagnosing the formation of Cooper pair condensates based on the Penrose-Onsager criterion, in which superconducting order is encoded in the spectral properties of the two-particle reduced density matrix (2RDM). Within this formulation, the symmetry and structure of the condensate are obtained by projecting the 2RDM onto irreducible representations of the underlying symmetry group, enabling an unbiased identification of both conventional and exotic superconducting states. We demonstrate the power and versatility of the approach through applications to the two-dimensional Hubbard model, using both auxiliary-field quantum Monte Carlo (AFQMC) and the density matrix renormalization group (DMRG). For attractive interactions without a magnetic field,  we reveal a clear finite-size scaling of the condensate fraction on square lattices of size up to $20\times 20$. The framework further provides direct access to the internal structure and extent of Cooper pairs, which we track across the BCS-BEC crossover.  Moreover, it enables a clean diagnosis of the finite-momentum Fulde-Ferrell-Larkin-Ovchinnikov (FFLO) phase in a magnetic field. Finally, we apply the approach to a supersolid phase in the repulsive Hubbard model with an additional next-nearest neighbor hopping $t^\prime$, where a charge-density wave coexists with a superconductor. We confirm the fragmented nature of the condensate and uncover substantial pairing correlations in the triplet channel with $p$-wave spatial symmetry in addition to the dominant singlet $d$-wave pairing. Our results establish the 2RDM-based Penrose-Onsager framework as a broadly applicable and unbiased tool for characterizing superconducting order in correlated quantum matter. 

\end{abstract}

\maketitle

\tableofcontents

\section{Introduction}
Understanding the origins of superconductivity is one of the prime objectives in condensed matter physics. In particular, superconductivity arising from electron-electron interaction alone poses significant challenges for theorists unraveling the intricacies of the interacting many-electron problem. In recent years, much progress has been made in understanding phase diagrams of strongly interacting Hubbard and $t$-$J$ models by certain numerical and analytical methods~\cite{chang10,zheng17,huang18,huang17,qin20,xu22,jiang24, xu24, baldelli_fragmented_2025, vanhala18, mai22,gros88, halboth00, maier05, senechal05, gull13}. The fundamental single-orbital Hubbard model on a two-dimensional square lattice has received particular attention due to its relevance to the high-temperature cuprate superconductors. Indeed, many aspects of the physics of cuprate superconductors are faithfully reproduced by this minimal model to a certain extent. Besides the antiferromagnetic Mott insulating regime at half-filling, stripe phases and superconducting regimes have now also been identified in the Hubbard model, and effects such as the pseudogap phenomenon have also been shown to be a consequence of the Hubbard model. Interestingly, the single-orbital Hubbard model also features supersolid phases, where charge density waves coexist with superconductivity. Beyond the cuprates, progress on models describing different strongly correlated superconductors, such as moiré systems, iron-based superconductors, nickelates, and kagom\'e metals, has been steep, where multiple superconducting phases have been identified~\cite{cao18,xia25,balents20,kamihara08,si16,fernandes22,zhou25,li25,li19,wang25,wu23,hossain25,liu24}. 

A key methodological question at the core of these studies is: how should one properly identify a superconducting state of matter? Quantum Monte Carlo and tensor network approaches have applied several approaches in the past. First, the detection of off-diagonal long-range order (ODLRO) is a well-established means of detecting spontaneous symmetry breaking. In the context of the density matrix renormalization group (DMRG), however, studies are often performed on cylinders. While typically long or even infinitely long cylinders can be studied, their width is typically limited to several lattice sites, ultimately rendering the geometry one-dimensional, prohibiting ODLRO. Instead, quasi-long-range order with algebraically decaying correlation functions is observed, often taken as an indicator of an emergent Luther-Emery liquid\cite{luther-emery}. A second approach is to slightly break the U($1$) particle number symmetry using a pinning field and studying the response of the system~\cite{jiang21,white09,venderley19,qin20}.

In this manuscript, we detail the application of the Penrose-Onsager criterion for establishing superconductivity. At the core of this method lies the two-particle reduced density matrix, its spectrum and eigenvectors. The leading eigenvalue is identified with the number of electrons in the condensate, for the Cooper-pair condensate. The corresponding eigenvector defines the pair-wave function describing the properties of the Cooper pairs. We apply this criterion to three distinct superconducting phases in the square-lattice Hubbard model. First, a $s$-wave superconducting state in the attractive, spin-balanced Hubbard model, which is studied using auxiliary-field quantum Monte Carlo and DMRG. We demonstrate the scaling of the leading eigenvalue with particle number and investigate the size of the Cooper pair across the BCS-BEC crossover~\cite{randeria14,zwerger12}. We then study the Fulde–Ferrell–Larkin–Ovchinnikov (FFLO) phase in the spin-imbalanced attractive Fermi-Hubbard model using DMRG and compare this state to the supersolid state in the repulsive Hubbard model at finite doping. Intriguingly, the latter state realizes a fragmented superconducting state. 

\section{Penrose-Onsager criterion}
The central object of study for superconductivity in strongly correlated electron systems is the two-particle reduced density matrix ($2$RDM) defined as,
\begin{equation}
\label{eq:full2rdm}
\rho_2(\br_{1}\sigma_{1}\;\br_{2}\sigma_{2}:\br_{1}'\sigma_{1}'\;\br_{2}'\sigma_{2}')=\langle c_{\br_1\sigma_{1}}^{\dagger} c_{\br_2\sigma_{2}}^{\dagger}c_{\br_2'\sigma_{2}'}c_{\br_1'\sigma_{1}'} \rangle,
\end{equation}
where $c^\dagger_{\br\sigma}$ ($c_{\br\sigma}$) creates (annihilates) an electron with spin $\sigma$ on site $\bm{r}$ defined on some lattice $\mathcal{L}$. Crucially, $\rho_2$ is Hermitian,
\begin{equation}
\rho_2(\br_{1}\sigma_{1}\;\br_{2}\sigma_{2}:\br_{1}'\sigma_{1}'\;\br_{2}'\sigma_{2}') = \rho^*_2(\br_{1}'\sigma_{1}'\br_{2}'\sigma_{2}':\br_{1}\sigma_{1}\br_{2}\sigma_{2}),
\end{equation}
and thus allows for a spectral decomposition, 
\begin{multline}
\rho_2(\br_{1}\sigma_{1}\;\br_{2}\sigma_{2}:\br_{1}'\sigma_{1}'\;\br_{2}'\sigma_{2}')  \\
= \sum_l \varepsilon_l \;\psi_l^*(\br_{1}\sigma_{1},\br_{2}\sigma_{2})\psi_l(\br_{1}'\sigma_{1}',\br_{2}'\sigma_{2}'),
\end{multline}
where $\varepsilon_l$ denotes the $l$-th (real) eigenvalue and $\psi_l(\br_{1}\sigma_{1},\br_{2}\sigma_{2})$ the corresponding normalized eigenvector. Notice, that in general the eigenvectors $\psi_l$ are functions of two positions $\br_{1}$, $\br_{2}$ and spins $\sigma_{1}$, $\sigma_{2}$. We thus interpret $\psi_l$ as pair wave functions. The scaling of the leading eigenvalues as a function of the number of electrons $N_e$ allows for three possible scenarios~\cite{Leggett}:
\begin{enumerate}
    \item \textit{Normal state:} all eigenvalues $\varepsilon_l$ are of order unity.
    \item \textit{Simple condensate:} exactly one eigenvalue $\varepsilon_0$ is of order $N_e$, the remaining eigenvalues are of order unity.
    \item \textit{Fragmented condensate:} two or more eigenvalues are of order $N_e$, the remaining eigenvalues are of order unity. 
\end{enumerate}
This criterion for the formation of condensates is known as the Penrose-Onsager criterion~\cite{penrose-onsager, Yang1962, Leggett2022}. The lesser-known fragmented condensates have been observed experimentally in spinor Bose-Einstein condensates and driven condensates~\cite{evrard21, luo17}. Interestingly, evidence for condensate fragmentation has also recently been reported for striped superconductors, or supersolids, in the repulsive Hubbard model~\cite{wietek22,baldelli_fragmented_2025} and bipolaronic superconductors in the Su-Schrieffer-Heeger model~\cite{grundner23}. 

For a simple condensate, the pair wave function $\psi_0$ corresponding to the dominant eigenvalue $\varepsilon_0$ is interpreted as the condensate wave function of the Cooper pairs. Furthermore, the superconducting order parameter is conventionally defined as~\cite{Leggett},
\begin{equation}
    F(\br_1\sigma_1,\br_2\sigma_2) = \sqrt{\varepsilon_0}\; \psi_0(\br_1\sigma_1, \br_2\sigma_2).
\end{equation}

The full 2RDM as defined in \cref{eq:full2rdm} is an object consisting of four spatial and spin coordinates, $\bm{r}_1\sigma_1, \bm{r}_2\sigma_2, \bm{r}_1^\prime\sigma_1^\prime, \bm{r}_2^\prime\sigma_2^\prime$. However, to diagnose the formation of a condensate it can be sufficient to investigate certain projected matrices. Let,
\begin{equation}
    \mathcal{L}_2 = \left\{ (\bm{r}_1\sigma_1, \bm{r}_2\sigma_2)\; | \; \bm{r}_1, \bm{r}_2 \in L, \; \sigma_1, \sigma_2 \in \{ \uparrow, \downarrow\} \right\}, 
\end{equation}
be the discrete set of pairs of combined lattice and spin coordinates. Then we define the vector space of complex-valued functions on $\mathcal{L}_2$,
\begin{equation}
\label{eq:labelspace}
    \mathcal{C} = \{\psi: \mathcal{L}_2\rightarrow \mathbb{C}\} = \mathbb{C}^{\mathcal{L}_2}.
\end{equation}
The 2RDM can now be interpreted as a linear operator $\rho_2$ acting on $\mathcal{C}$. Let us denote by $P$ a projection matrix on $\mathcal{C}$ satisfying $P^2 = P$. Then, according to the Rayleigh-Ritz principle,
\begin{equation}
    \label{eq:poincareseparation}
    \varepsilon_0 = 
    \max\limits_{v\in \mathcal{C}}\frac{v^\dagger \,\rho_2 \, v}{v^\dagger v} \geq 
    \max\limits_{v\in P(\mathcal{C})}\frac{v^\dagger \,\rho_2 \, v}{v^\dagger v} = \varepsilon_0^P,
\end{equation}
where $P(\mathcal{C}) \subseteq \mathcal{C}$ denotes the projected space under $P$ and $\varepsilon_0^P$ denotes the largest eigenvalue of the projected 2RDM, $P^\dagger\rho_2 P$. \Cref{eq:poincareseparation}
is also known as the \textit{Poincar\'e separation theorem}. Thus, whenever the eigenvalue $\varepsilon_0^P$ of a projected 2RDM scales linearly in the number of electrons, this implies that $\varepsilon_0 \geq \varepsilon_0^P$ necessarily also needs to scale (at least) linearly in the number of electrons. In other words, the formation of a condensate in a projected 2RDM implies the formation of a condensate in the full 2RDM. We will discuss several instances, such as momentum or singlet projections, in the following sections. 

The 2RDM in \cref{eq:full2rdm} is defined with real-space creation and annihilation operators $c^\dagger_{\br \sigma}$ and $c_{\br \sigma}$. With periodic boundary conditions, we can equivalently also define the 2RDM in momentum space, 
\begin{equation}
\label{eq:full2rdmkspace}
\rho_2(\bk_{1}\sigma_{1}\;\bk_{2}\sigma_{2}:\bk_{1}'\sigma_{1}'\;\bk_{2}'\sigma_{2}')=\langle c_{\bk_1\sigma_{1}}^{\dagger} c_{\bk_2\sigma_{2}}^{\dagger}c_{\bk_2'\sigma_{2}'}c_{\bk_1'\sigma_{1}'} \rangle,
\end{equation}
where annihilation operator in momentum space is defined as, 
\begin{equation}
    c_{\bk \sigma} = \frac{1}{\sqrt N}\sum_{\br} e ^{-i\bk \cdot \br} c_{\br \sigma},
\end{equation}
where $N$ denotes the number of lattice sites. This definition corresponds to a change of basis of $\mathcal{C}$,
\begin{multline}
\rho_2(\bk_{1}\sigma_{1}\;\bk_{2}\sigma_{2}:\bk_{1}'\sigma_{1}'\;\bk_{2}'\sigma_{2}') \\ = U^\dagger \rho_2(\br_{1}\sigma_{1}\;\br_{2}\sigma_{2}:\br_{1}'\sigma_{1}'\;\br_{2}'\sigma_{2}') U,
\end{multline}
where this unitary similarity transformation is implemented via the Fourier transform,
\begin{equation}
    (U)_{\bk_{1}\sigma_1\bk_{2}\sigma_2, \br_{1}\sigma_1\br_{2}\sigma_2} = \frac{1}{N} e^{-i\bk_{1}\cdot \br_{1}} e^{-i\bk_{2} \cdot\br_{2}}. 
\end{equation}
The spectrum of any matrix is invariant under a unitary similarity transformation. Thus, the spectra of the 2RDM in real space and momentum space will coincide and investigations concerning the nature of condensates can be performed in both bases.

\section{Symmetries of the condensate}

To characterize symmetry properties of a condensate, we consider projectors onto irreducible representations (irrep) of the symmetry group.
For simplicity, let us consider a discrete symmetry group $G$ and denote by, 
\begin{equation}
 \pi: G \rightarrow \text{GL}(\mathcal{C}),   
\end{equation}
the action of $G$ on the space $\mathcal{C}$ as in \cref{eq:labelspace}. We define the projector onto an irrep $\alpha$ as,
\begin{equation}\label{eq:projector_irrep}
    P_\alpha = \frac{d_\alpha}{|G|}\sum_{g\in G}\chi^*_\alpha(g) \pi(g),
\end{equation}
where $d_\alpha$ denotes the dimension of $\alpha$, and $\chi_\alpha(g)$ denotes the character of $\alpha$ evaluated at $g$. This definition gives rise to the notion of the "symmetry" of the condensate. 

We define an $\alpha$-condensate to be a condensate where the leading eigenvalue of the $\alpha$-projected 2RDM, 
\begin{equation}
    \rho_2^\alpha = P_\alpha^\dagger \rho_2 P_\alpha,
\end{equation}
is also of order $N_e$. Moreover, assume an eigenvector $\psi_l\in \mathcal{C}$ with eigenvalue $\varepsilon_l$ of the full 2RDM $\rho_2$ transforms according to a one-dimensional irrep $\alpha$ of $G$, 
\begin{equation}
\label{eq:psitransformirrep}
    \pi(g)\psi_l = \chi_\alpha(g)\psi_l.
\end{equation}
Then $\psi_l$ is invariant under projection, $P_\alpha \psi_l = \psi_l$, and remains an eigenvector of the projected 2RDM, again with eigenvalue $\varepsilon_l$,
\begin{equation}
    \rho^\alpha_2 \psi_l = (P_\alpha^\dagger \rho_2 P_\alpha) \psi_l = \varepsilon_l \psi_l.
\end{equation}
Specifically, consider the case $l=0$ for a simple condensate and $\psi_0$ transforms with irrep $\alpha$ according to \cref{eq:psitransformirrep}. Then, $\varepsilon_0$ is also an eigenvalue of the projected 2RDM $\rho_2^\alpha$ scaling linearly with the number of electron $N_e$, which implies that the condensate is an $\alpha$-condensate.

\begin{figure}[ht]
    \centering
    \includegraphics[width=\linewidth]{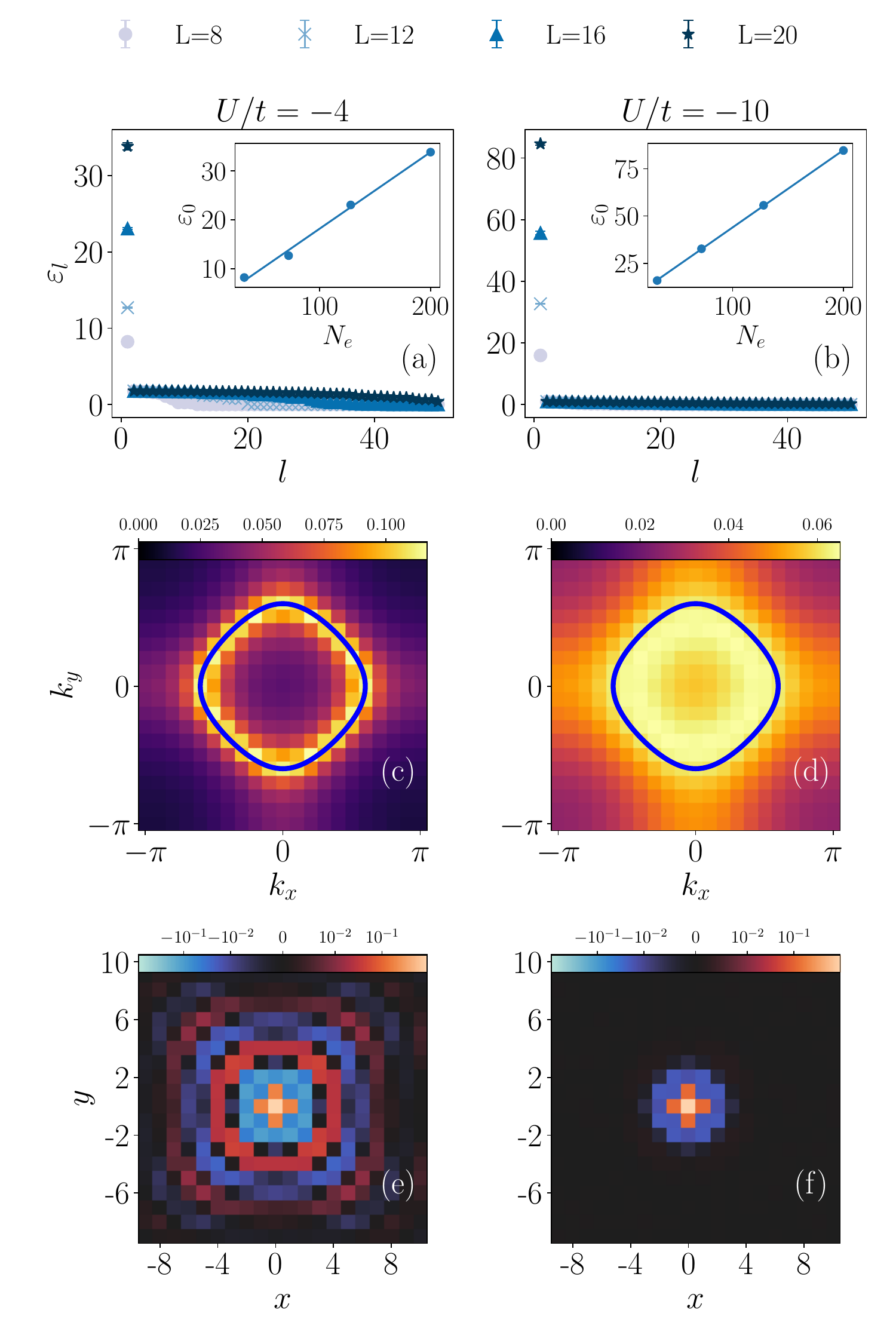}
    \caption{Spectral properties of the $2$RDM and the leading eigenvector for the two-dimensional attractive Hubbard model at density $n=0.5$, for $U/t=-4$ (a,c,e) and $U/t=-10$ (b,d,f) obtained using AFQMC. (a, b) Spectrum of the $2$RDM. For both $U/t=-4$ and $U/t=-10$ we see a dominant eigenvalue growing linearly in the number of electrons $N_e$. (c, d) The condensate wave function in momentum space for $L=20$. For $U/t=-4$, the condensate wave function peaks around the non-interacting Fermi surface, indicated by the blue line. For $U/t=-10$, the wave function peaks slightly inside of the Fermi surface. (e, f) The condensate wave function in real space for $L=20$ is invariant under $C_4$ rotations. These observations establish the formation of a Cooper condensate with uniform $s$-wave pairing.}
    \label{fig:afqmc_wave function}
\end{figure}

Given a complete set $\{\alpha\}$ of irreducible representations of the symmetry group $G$, the $2$RDM can be fully decomposed into irrep contributions,
\begin{equation}
    \rho_2 = \sum_\alpha \rho^\alpha_2.
\end{equation}
These definitions allow us to describe properties like the momentum of the condensate or its transformation properties under point group operations. We will discuss specific examples of symmetry in condensates in the following section.

\section{Superconductivity in the Hubbard model}
In order to study properties of distinct superconducting states, the focus of our attention will be the two-dimensional $t$-$t^\prime$-$U$ Hubbard model on the square lattice, given by,
\begin{multline}
    H = 
    -t\sum_{\braket{ij},\sigma} (c_{i\sigma}^\dagger c_{j\sigma} + \text{h.c.}) \\
    -t^\prime\sum_{\braket{\braket{ij}},\sigma} (c_{i\sigma}^\dagger c_{j\sigma} + \text{h.c.})+U\sum_i n_{i\uparrow}n_{i\downarrow}. 
\end{multline}
where $c_{i\sigma}^\dagger$ and $c_{i\sigma}$ denote the creation (annihilation) operators for electrons on sites $\br_i$ and spin $\sigma$, $\braket{i,j}$ denotes summation over nearest-neighbor lattice sites, and $\braket{\braket{i,j}}$ denotes the diagonal next-nearest-neighbors.

In the attractive, $U/t<0$, regime, with $t'/t=0$ at magnetization $M=0$, the ground state has been shown to be a singlet s-wave superconductor with net momentum $\bm{q}=0$, for any finite interaction $U$~\cite{lieb89,scalettar89,moreo91,singer98,superconductivitylowenergyexcitations}, as predicted by BCS theory~\cite{BCS,waldram96}. For magnetization $M\neq 0$, the existence of FFLO states has also been shown, as well as a Fermi-liquid phase~\cite{luscher08,parish07,cheng18}.

The repulsive, $U/t>0$, regime exhibits a significantly richer phase diagram, as a function of doping, temperature, and values of $U/t$ and $t^\prime/t$. At half-filling, $n=1$, a Mott insulating antiferromagnetic state has been clearly established. At finite doping, the occurrence of several competing phases has been widely discussed. Recent studies point towards distinct stripe states being stabilized as a ground state at low hole-doping~\cite{chang10,zheng17,huang18,huang17,qin20,xu22}. Moreover, there is ample evidence that non-zero values of $t^\prime/t\neq 0$ enhance pairing correlations, likely leading to a superconducting state. The superconducting state at low doping has often been reported as coexisting with a charge density wave (CDW)~\cite{jiang24, xu24, baldelli_fragmented_2025, vanhala18, mai22}, whereas at higher doping, a uniform superconducting state without CDW order is observed~\cite{gros88, halboth00, maier05, senechal05, gull13}.

\subsection{Uniform $s$-wave in two dimensions}
\begin{figure*}
    \centering
    \includegraphics[width=\linewidth]{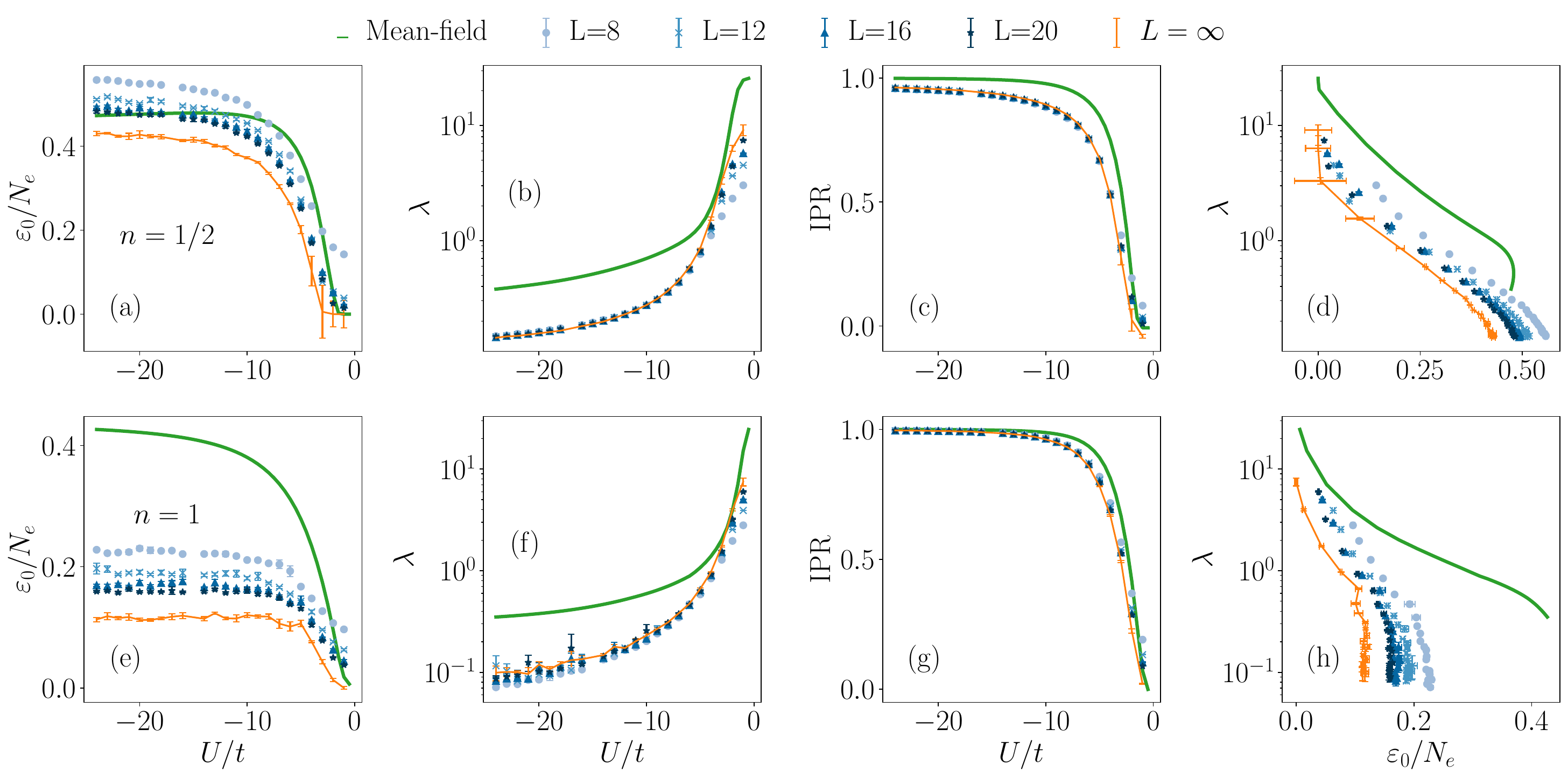}
    \caption{Condensation and localization properties of Cooper pairs in the two-dimensional attractive Hubbard model. We compare results from AFQMC and BCS mean-field theory for quarter-filling $n=1/2$ in the upper row (a-d) and half-filling $n=1$ in the lower row (e-h). The orange lines for $L=\infty$ are obtained using AFQMC by extrapolation in $L$. (a,e) Condensate fraction $\varepsilon_0 / N_e$, (b,f) localization length $\lambda$, (c,g) inverse participation ratio IPR, as a function of $U/t$. (d,h) Scaling of the localization length with the condensate fraction.   We observe a transition from delocalized Cooper pairs at small values of $U/t$ to fully localized Cooper pairs for larger values of $U/t$.
}
\label{fig:afqmc_data}
\end{figure*}

We begin our discussion with the case of the superconducting state in the attractive Hubbard model, $U/t<0$ and $t'/t=0$, at magnetization $M=0$. To this end, we investigate the singlet 2RDM with momentum $\bm{q}$,
\begin{equation}
    \rho_2^{\bm{q},S}(\bk, \bk^\prime) = \braket{\Delta^\dagger_{\bk, \bm{q} - \bk} \Delta_{\bk^\prime, \bm{q} - \bm{k}^\prime}},
\end{equation}
where,
\begin{equation}
\Delta_{\bk_1\bk_2}^\dagger = \frac{1}{\sqrt{2}}(c^\dagger_{\bk_1\uparrow}c^\dagger_{\bk_2\downarrow} - c^\dagger_{\bk_1\downarrow}c^\dagger_{\bk_2\uparrow}).
\end{equation}
The matrix $\rho_2^{\bm{q},S}(\bk, \bk^\prime)$ can be viewed as a projected 2RDM, where the projection operator is explicitly given as,
\begin{align}
    P_{\bm{q},S} &= \delta_{\bk_1 + \bk_2, \bm{q}}
    \otimes P_S, \\ 
    P_S &= \frac{1}{4}\left( I - \vec{\sigma}_1 \cdot \vec{\sigma}_2 \right). 
    \label{eq:singletprojector}
\end{align}
Here, $P_S$ denotes the singlet projection operator and $\vec{\sigma} = (\sigma^x, \sigma^y, \sigma^z)$ denote the Pauli matrices. Technically, the projected matrix matrix $P_{\bm{q},S}^\dagger \rho_2P_{\bm{q},S}$ is still a function of four momentum and spin variables $\bk_{1}\sigma_{1}, \bk_{2}\sigma_{2},\bk_{1}'\sigma_{1}',\bk_{2}'\sigma_{2}'$.
The matrix $\rho_2^{\bm{q},S}(\bk_1, \bk_2)$ can be considered a marginalized density matrix, i.e.
\begin{equation}
    \rho_2^{\bm{q},S}(\bk_1, \bk^\prime_1) = \text{Tr}_{\bk_{2}\bk_{2}^\prime}\text{Tr}_{\sigma_1\sigma_2\sigma_1^\prime\sigma_2^\prime}\left[P_{\bm{q},S}^\dagger \;\rho_2\; P_{\bm{q},S}\right].
\end{equation}
Moreover, we focus on the total momentum zero case with $\bm{q} =0$. We consider the eigendecomposition,
\begin{equation}
 \rho_2^{\bm{q}=0,S} = \sum_l \varepsilon_l \psi_l(\bk)^* \psi_l(\bk),
\end{equation}
from which we obtain the pair wave functions $\psi_l(\bk)$ in momentum space, which are a function of a single momentum $\bk$.

In order to numerically compute $\rho_2^{\bm{q}=0,S}$, we employ the projective auxiliary-field quantum Monte Carlo method~\cite{blankenbecler81}. 
The ground state of the interacting Hamiltonian is determined by imaginary-time evolution of a suitable trial wave function,
$|\Phi\rangle_{\mathrm{GS}}=\lim_{\theta\rightarrow\infty}e^{-\theta H}|\Phi\rangle_{\mathrm{trial}}$, within a given particle number sector. The trial wave function is a Slater determinant of $N_e$ electrons that preserves time-reversal symmetry to maintain the absence of the sign-problem and a non-zero overlap $\langle \Phi_{\mathrm{GS}} | \Phi_{\mathrm{trial}}\rangle \neq 0$ ensures convergence for large $\theta$. We point out that $\rho_2^{\bm{q}=0,S}$ can be readily obtained~\cite{shi16,rosenberg17} and is related to the traditionally used s-wave singlet pair correlation function. For most set of parameters we use a projection time of $\theta=20$, for which we confirm convergence, see details in Appendix~\ref{app:afqmc}. For some more difficult parameters at weak coupling and lower density, however, like $n=0.5$, $L=20$, and $U=-1$, we use a projection time of up to $\theta=200$. Our simulations are performed using the \textit{ALF: Algorithms for Lattice Fermions} library~\cite{ALF}.

Our first goal is to demonstrate the formation of a condensate, for which we investigate the scaling of the leading eigenvalue, $\varepsilon_0$, of the 2RDM $\rho^{\bm{q}=0,S}_2$ as a function of system size. For a fixed filling, this is equivalent to studying how the leading eigenvalue scales with the number of electrons, $N_e$. We focus on quarter-filling, i.e., $n=N_e/N = 1/2$, and compare the results of linear system sizes $L=8,12,16,20$ in \cref{fig:afqmc_wave function} (a,b) for $U/t=-4$ and $U/t=-10$, respectively. The dominant eigenvalue is interpreted as the number of electrons in the condensate, which we observe to grow linearly in the number of electrons. According to the Penrose-Onsager criterion, the system is thus a superconductor. The corresponding eigenvector, interpreted as the condensate wave function in momentum space, is shown in~\cref{fig:afqmc_wave function}(c,d). For $U=-4$, the condensate wave function peaks at values close to the Fermi surface of the non-interacting system, as expected from BCS theory~\cite{BCS,annett04}. The wave function is more spread out as the magnitude of the interaction strength is increased to $U/t=-10$, and now peaks slightly inside the Fermi surface rather than around it.
We now define the condensate wave function in real space by, 
\begin{equation}
    \psi_0(\br) = \frac{1}{\sqrt{N}}\sum_{\bk} e^{i\bk \cdot \br} \psi_0(\bk),
\end{equation}
and the results are shown in \cref{fig:afqmc_wave function}(e,f). We observe oscillations in sign with the distance between the electrons, as well as a decrease in magnitude. Here, $\br$ is viewed as the relative distance between the two electrons in the Cooper pair. As we increase the interaction strength, the Cooper pair gets more localized in real space, while the region of pairing in momentum space grows.
In order to measure the size of the Cooper pair, we investigate the localization length $\lambda$ given by,
\begin{equation}
    \lambda^2 = \sum_{\br} \lVert\br\lVert^2 |\psi_0(\br)|^2,
\end{equation}
where $\lVert \br \lVert$ denotes the two-norm of $\br$. We study the dependence of $\lambda$ on both the interaction strength as well as the condensate fraction $\varepsilon_0/N_e$.
We consider the cases of quarter-filling, $n=1/2$, and half-filling, $n=1$, in \cref{fig:afqmc_data} (a, c, e, g). Moreover, we consider the inverse participation ratio (IPR) of the Cooper pair defined as,
\begin{equation}
    \text{IPR}= \sum_{\br} |\psi_0( \br)|^4.
\end{equation}
The IPR measures how distributed the Cooper pair is in real space, rather than the size of the Cooper pair, as for the localization length. If the pair is fully localized, the IPR yields 1. If it is fully delocalized, the IPR yields $\sum_{\br} (1/\sqrt{N})^4=1/N^2$.

We compare our results to BCS mean-field theory in \cref{fig:afqmc_data}, see Appendix~\ref{app:mf} for details concerning the mean-field theory. We find that our results for $\lambda$ and the IPR from AFQMC extrapolated to $L=\infty$ are reasonably well described by mean-field theory, qualitatively. Since our data is consistent with $\lambda$ diverging as $U/t\to 0$, and the IPR rapidly decreases, the Cooper pair becomes delocalized in this limit. Intuitively, at small values of $U/t$ the pairing is localized in momentum space in a narrow ribbon around the Fermi surface, causing it to delocalize in real space. Thus, the localization length $\lambda$ should diverge, and the IPR should behave as $1/N^2$. Concerning the dependence of $\varepsilon_0$ on $U/t$, we initially observe a rapid increase in the dominant eigenvalue as we increase the interaction strength. For larger values of $|U/t|$, although $\varepsilon_0/N_e$ saturates to a finite value $\leq 1$, we can see from panels (d,h) that the Cooper pair continues to further localize upon increasing $U/t$.

\subsection{Uniform $s$-wave on a cylinder}

\begin{figure}
    \centering
    \includegraphics[width=\linewidth]{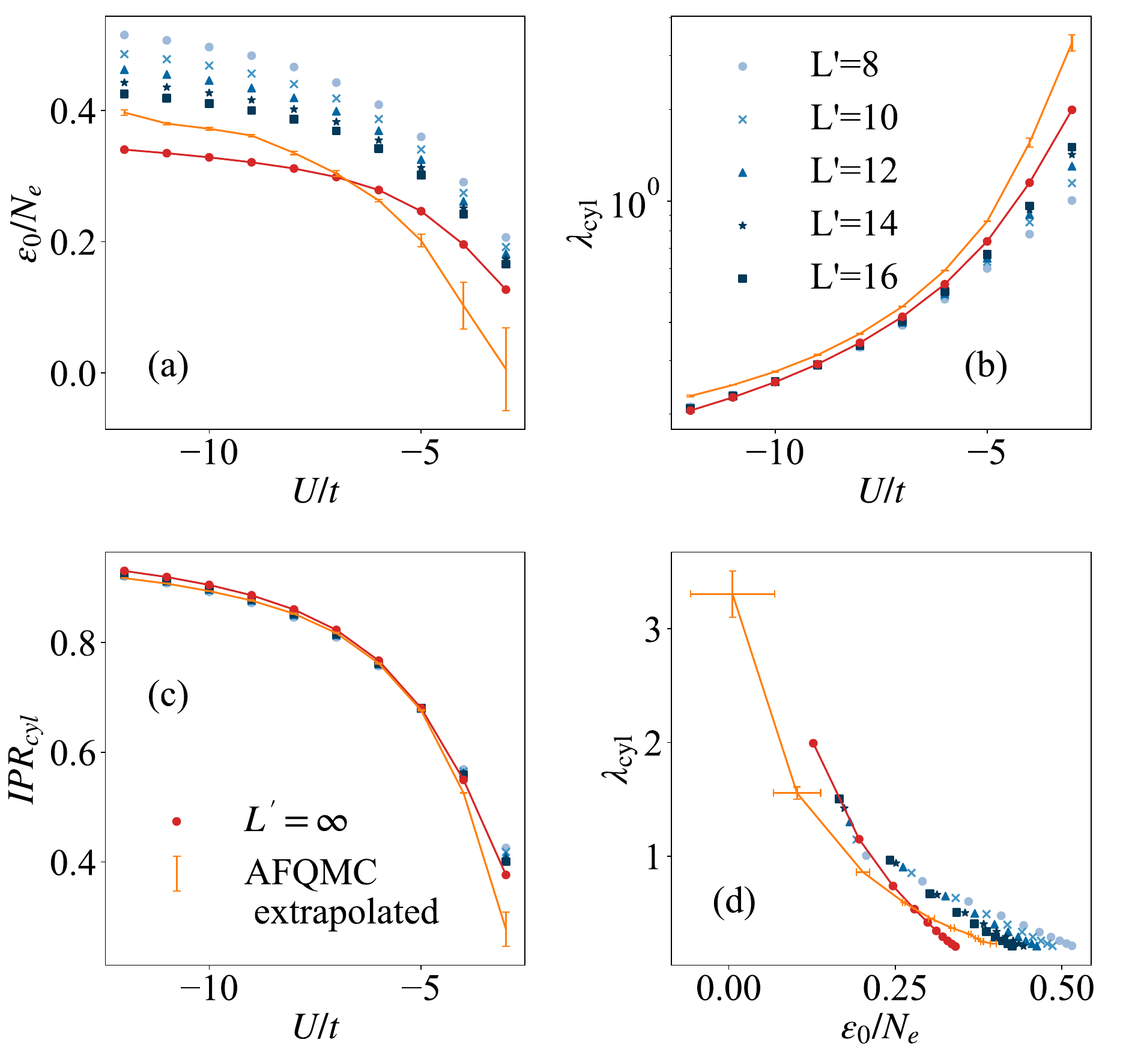}
    \caption{Condensation and localization properties of Cooper pairs in the two-dimensional attractive Hubbard model on a cylinder of width $W=4$. We show results from DMRG, for quarter-filling $n=1/2$ and compare with the results from AFQMC (orange) on $L\times L$ lattices, extrapolated to $L=\infty$. The red lines for $L=\infty$ are obtained using DMRG by extrapolation in $L^\prime$. (a) Condensate fraction $\varepsilon_0/N_e^\nu$, (b) localization length $\lambda_{\text{cyl}}$, (c) Inverse participation ratio $\text{IPR}_{\text{cyl}}$ as a function of $U/t$. (d) Scaling of the localization length with condensate fraction.}
    \label{fig:dmrg}
\end{figure}

\begin{figure}
    \centering
    \includegraphics[width=\linewidth]{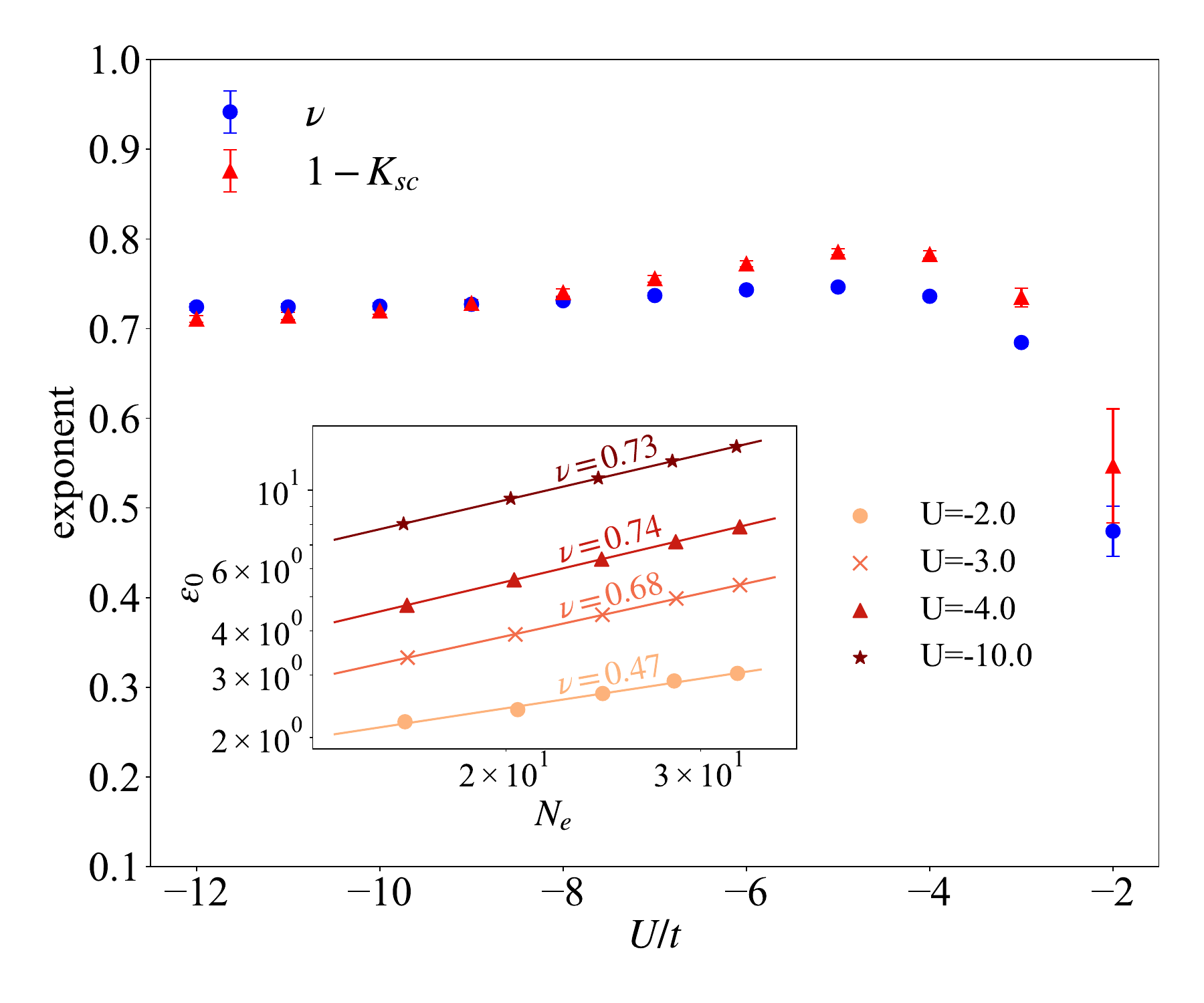}
    \caption{Comparison between the exponents $\nu$ and $1-K_{sc}$ as functions of $U$, where $K_{sc}$ is the superconducting Luttinger parameter. The inset displays the dominant eigenvalue $\varepsilon_0$ as a function of the number of electrons, $N_e$, for a range of values of interaction strength $U$, and shows the exponent $\nu$ with which it grows. We observe a good agreement between $\nu$ and $1-K_{sc}$, and see that the dominant eigenvalues grows algebraically, but sub-linearly, in $N_e$.}
    \label{fig:exponent}
\end{figure}

One of the drawbacks of QMC methods is the well-known sign problem~\cite{ALF}, making it difficult to simulate e.g., fermions with repulsive interactions or spin-imbalanced systems. One then has to use different methods, of which a popular choice is DMRG~\cite{white92, white93}, together with matrix product states (MPS)~\cite{schollwock11}. However, DMRG comes with its own drawbacks, e.g., it is not efficient for higher dimensional systems, since the structure of the MPS is one-dimensional. To keep the computational costs reasonable, we constrain the system to a cylinder geometry, where we have periodic boundary conditions in the smaller $y$-direction, and open boundary conditions in the larger $x$-direction. However, the cylinder geometry is quasi one-dimensional, and hence the  Mermin-Wagner theorem prohibits genuine long range order~\cite{mermin-wagner, hohenberg}. Therefore, one has to take special care when comparing results on a cylinder with two-dimensional lattices.

We consider the singlet 2RDM in real space defined as, 
\begin{align}
\label{eq:singletpairingmatrix}
\rho^S_2(\bm{r}_1 \bm{r}_2 : \bm{r}_1^\prime \bm{r}_2^\prime) &=\text{Tr}_{\sigma_1\sigma_2\sigma_1^\prime\sigma_2^\prime}\left[P_S^\dagger \rho_2 P_S \right]\\
&=\langle \Delta_{\bm{r}_1\bm{r}_2}^\dagger \Delta_{\bm{r}_1^\prime\bm{r}_2^\prime}\rangle
\end{align}
where the singlet-pairing operators $\Delta^\dagger_{\bm{r}_i\bm{r}_j}$ are now defined as
\begin{equation}
\label{eq:singletpairingop}
\Delta_{\bm{r}_1\bm{r}_2}^\dagger= \frac{1}{\sqrt{2}}
\left( 
c^\dagger_{\br_1\uparrow} c^\dagger_{\br_2\downarrow} - 
c^\dagger_{\br_1\downarrow} c^\dagger_{\br_2\uparrow}
\right).
\end{equation}
Here, we project the 2RDM onto the singlet sector with $P_S$ as given in \cref{eq:singletprojector}. The DMRG simulations are performed with the ITensor package~\cite{itensor,itensor-r0.3}, on a cylinder at quarter-filling, $n=1/2$, with length $L=32$, and widths $W=1,2,4$. The results are obtained with bond dimensions of up to $D=4000$. To mitigate boundary effects, we only consider the $2$RDM for a subsystem at the center of the lattice, of size $L'\times W$, with $L^\prime<L$. The computation of the $2$RDM is made with the ITensorCorrelators~\cite{ITensorCorrelators} package. 

Since the condensate wave function, $\psi_0(\br_1,\br_2)$ now genuinely depends on two positions $\br_1=(x_1, y_1)$ and $\br_2=(x_2, y_2)$, we redefine the localization length $\lambda$ and the IPR. We choose a reference column in the middle of the cylinder, labeled as $x_1 = 0$, and define 
\begin{equation}
    \lambda^2_\text{cyl} = \frac{1}{W}\sum_{y_1,x_2,y_2} \lVert \br_2 - \br_1\lVert^2 \, |\psi_0(0,y_1,x_2,y_2)|^2,
\end{equation}
and similarly,
\begin{equation}
    \text{IPR}_\text{cyl} = \frac{1}{W}\sum_{y_1,x_2,y_2}|\psi_0(0,y_1,x_2,y_2)|^4,
\end{equation}
Qualitatively, the DMRG results agree with the AFQMC results, and the computed values are of the same order as the AFQMC results, see \cref{fig:dmrg}. We find however that the dominant eigenvalue no longer grows linearly in system size, but rather with a characteristic exponent $\nu$, such that $\varepsilon_0\sim N_e^\nu$. As a result, the cylinder geometry overestimates the magnitude of the dominant eigenvalue for low interaction strengths, and underestimate for large interaction strengths. Using the results for different values of $L'$, and thus different values of $N_e$, we can find $\nu$ through a linear fit $\log(\varepsilon_0) = a + \nu\log(N_e)$. The fits are shown in~\cref{fig:exponent}. We find that for $W=4$, the exponent is $\nu\approx 0.74$ for $U/t\leq -4$. For $W=1$ and $W=2$, the exponent $\nu$ is found to be $\nu\approx 0.39$, and $\nu\approx 0.51$ respectively. For $W=4$, the fit is made with $L'=8,10,12,14,16$, and for $W=1,2$, we use $L'=16,20,24,28,32$. The exponent $\nu$ is clearly increasing with width $W$, and should approach $\nu=1$ in the two dimensional limit, as with the AFQMC results. 
As for $\lambda$ and the IPR, see~\cref{fig:dmrg} (b, c), the DMRG and AFQMC results match well, with the AFQMC results being slightly larger for $\lambda$. This can likely be attributed to the finite width of the cylinder. 

We show in~\cref{sec:circulant} that the exponent $\nu$ is related to the superconducting Luttinger parameter $K_{sc}$ by $\nu=1-K_{sc}$. $K_{sc}$ is defined by the pair correlator $\braket{\Delta^\dagger(x) \Delta (0)} \sim x^{-K_{sc}}$, where $\Delta^\dagger(x)$ creates a Cooper pair at $x$. The values of $\nu$ found above give values of $K_{sc}=0.61$, $K_{sc}=0.49$, and $K_{sc}=0.25$ for $W=1$, $W=2$, and $W=4$ respectively. This is consistent with a Luther-Emery liquid~\cite{luther-emery}. For Luther-Emery liquids, is has been shown from analytical and DMRG studies that $K_{sc} \propto 1/W$ for large enough $W$~\cite{gannot23, jiang23,chatterjee22}, which we observe to hold well already for $W=2$ and $W=4$. We also compute $1-K_{sc}$ by fitting the pair correlator to a power law, and compare this with $\nu$ in~\cref{fig:exponent} (b).

\subsection{The Fulde-Ferrell-Larkin-Ovchinnikov state}

\begin{figure}
    \centering
    \includegraphics[width=\linewidth]{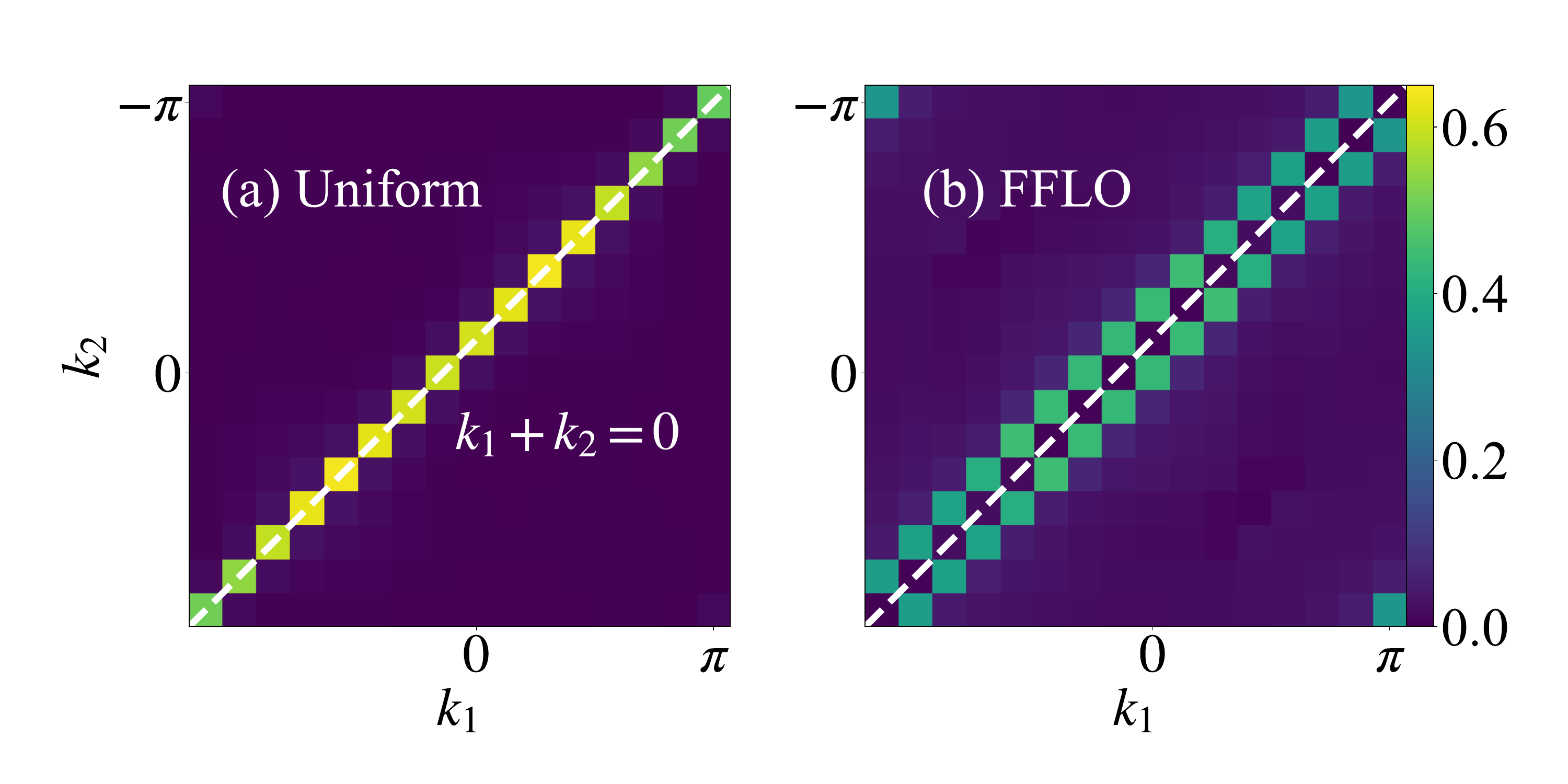}
    \caption{The condensate wave function $\tilde{\psi}_0(k_1, k_2)$ as function of momenta, $k_1$ and $k_2$, in the $x$-direction for the two electrons, at $U/t=-10$, quarter-filling $n=1/2$ and at magnetization (a) $m=0$ and (b) $m=1/32$. For $m=0$, the condensate wave function peaks at zero total momentum, but when introducing a spin imbalance, $m=1/32$, it peaks at non-zero momentum, indicating the existence of an FFLO state of LO type.}
    \label{fig:fflo}
\end{figure}

\begin{figure}
    \centering
    \includegraphics[width=\linewidth]{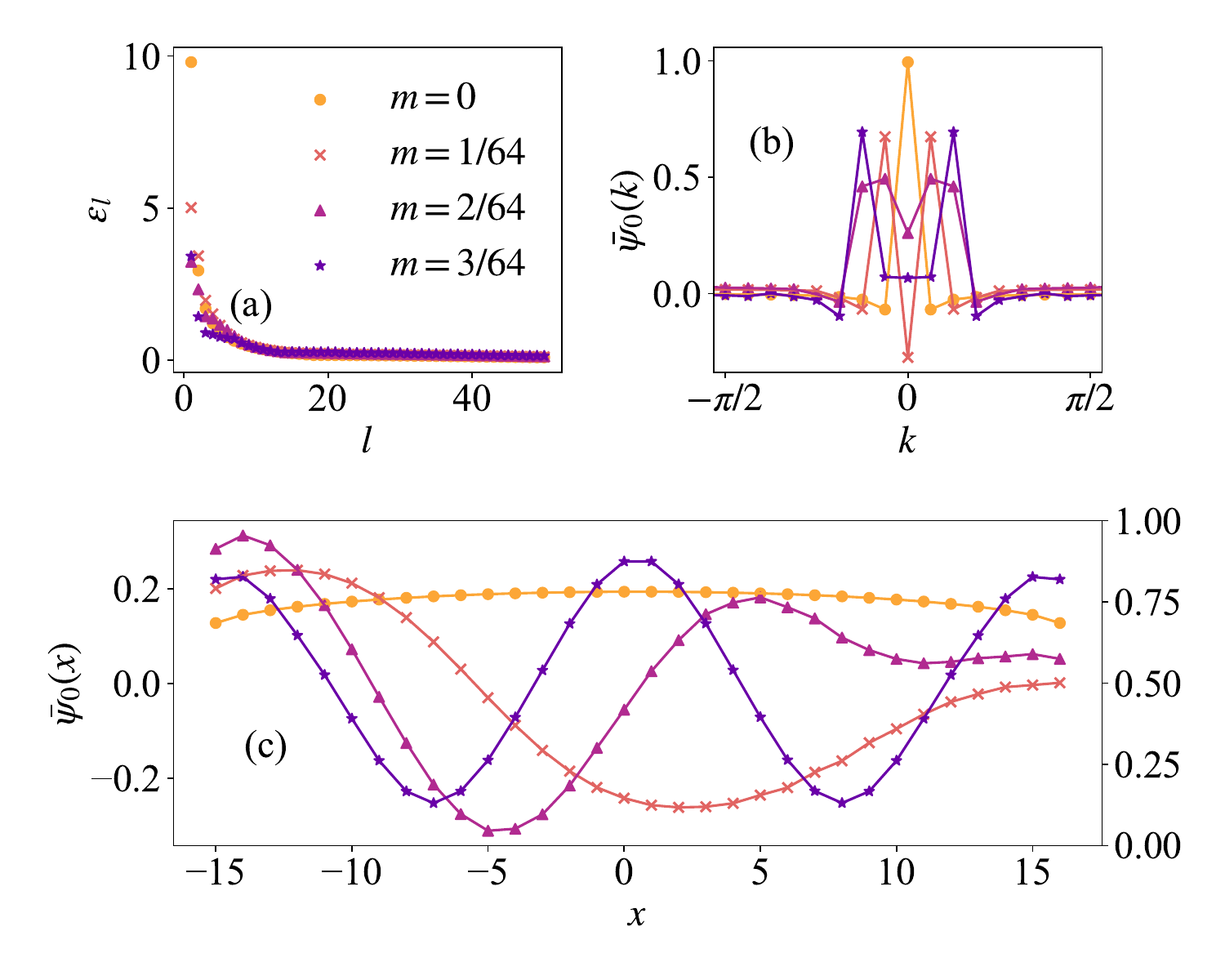}
    \caption{Condensate wave functions for the attractive Hubbard model at $U=-10$, quarter-filling $n=1/2$, at different magnetization on a $32\times4$ square lattice. (a) Spectrum of the local pairing $2$RDM. (b) Condensate wave functions in momentum space. (c) Condensate wave functions in momentum space. We observe that the dominant eigenvalue is decreasing as the magnetization is increased, and that the condensate wave functions develop periodic oscillations, consistent with FFLO states of LO type.}
    \label{fig:fflo_momentum}
\end{figure}

\begin{figure}
    \centering
    \includegraphics[width=\linewidth]{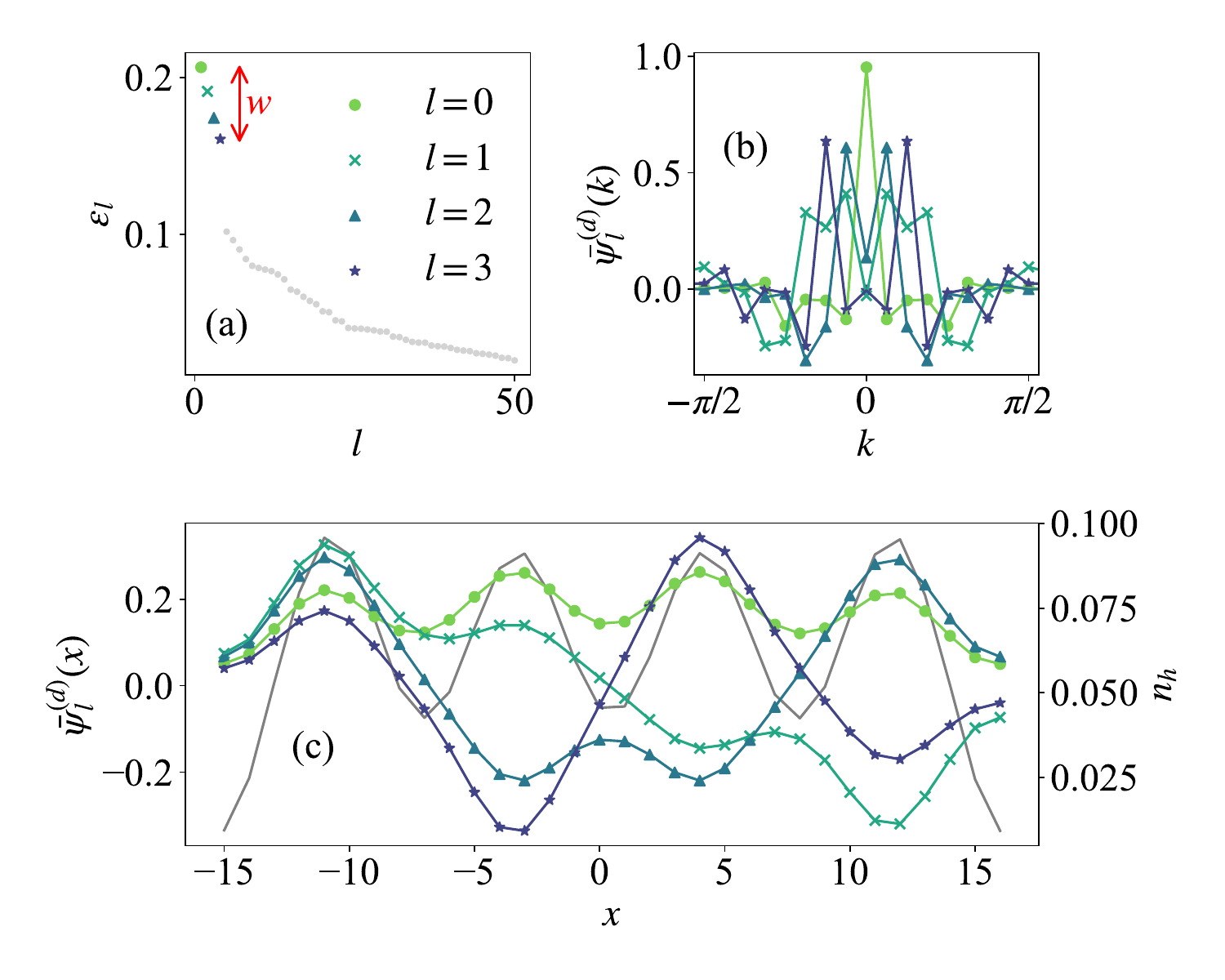}
    \caption{Condensate wave functions corresponding to fragmented condensates in the repulsive Hubbard model with $t'/t=-0.2$ on a $32\times4$ square lattice. (a) Spectrum of the nearest-neighbour pairing $2$RDM. Bandwidth $w$ shown in red. (b) Condensate wave functions in momentum space. (c) Condensate wave functions in real space, together with the hole density $n_h$ in grey. We observe that the different condensates display periodic oscillations corresponding to different momenta.}
    \label{fig:fragmented_momentum}
\end{figure}

We now turn to the case of a spin-imbalanced system, $M\neq 0$, which poses significant challenges for simulation using QMC methods due to the sign problem. When we introduce a spin-imbalance or magnetic field to a superconducting system, we expect it to either lose its superconducting properties, or develop a so-called FFLO state~\cite{Fulde-Ferrell, Larkin-Ovchinnikov}. There exist two versions of the FFLO state: the Fulde-Ferrell state (FF)\cite{Fulde-Ferrell}, where the condensate wave function carries finite momentum $\bm q$. Hence, the wave function transforms under a translation by $\bm{R}$ as
\begin{equation}
    \psi_0(\bm{r_1} + \bm{R}, \bm{r_2} + \bm{R}) = e^{i\bm q \cdot \bm R} \, \psi_0(\bm r_1, \bm r_2),
\end{equation}
which in general breaks time reversal symmetry, $\bm{q} = \rightarrow -\bm q$. The Larkin-Ovchinnikov state (LO)~\cite{Larkin-Ovchinnikov} on the other hand, is a superposition of two wave functions at finite momenta $\bm q$ and $-\bm q$, for example,
\begin{equation}
    \begin{split}
        \psi_0(\br_1, \br_2) &\propto e^{i\bm q\cdot (\br_1 + \br_2)/2} + e^{-i\bm q\cdot (\br_1 + \br_2)/2} \\
    &= 2 \cos[\bm q \cdot (\br_1 + \br_2)/2].
    \end{split}
\end{equation}
These states are understood to be instabilities of momentum-imbalanced Fermi surfaces. In one spatial dimension, the momentum of the Cooper pairs is given by~\cite{cheng18, feiguin07}, 
\begin{equation}
    q=|k_{F\uparrow}-k_{F\downarrow}|,
\end{equation}
where $k_{F\sigma}$ is the Fermi momentum for electrons with spin $\sigma$. Apart from being studied in electronic systems with magnetic fields, FFLO states are also of interest in heavy-fermion systems~\cite{bianchi03, matsuda07}, organic superconductors~\cite{shimahara08,imajo21, wosnitza18}, quantum chromodynamics, nuclear physics, and astrophysics~\cite{casalbuoni04}.

For the attractive Hubbard model, we perform DMRG simulations on a $32 \times 4$ cylinder at quarter-filling, $n=1/2$, for a range of values of interaction strength $U$ and magnetization density, 
\begin{equation}
m= \frac{1}{2N}(N_\uparrow-N_\downarrow),    
\end{equation}
where $N_\sigma$ denotes the electron number with spin $\sigma$. Due to higher entanglement, the maximum bond dimension was chosen to be $D=6000$. As in the previous section, we compute $\rho_2^S$, and find the condensate wave function $\psi_0$ by diagonalization. Due to the open boundary conditions, we lack translational invariance and momentum is thus not a good quantum number. Despite this, we can still make a Fourier analysis of the momentum content in the wave function. Summing over the $y$-dependence of the order parameter and then taking the Fourier transform
\begin{equation}
    \tilde{\psi}_0(k_1,k_2) = \sum_{x_1,x_2} e^{-i(k_1x_1 + k_2x_2)}\sum_{y_1,y_2} \psi_0(x_1,y_1,x_2,y_2) ,
\end{equation}
we get a momentum space representation in the $x$-direction. The case of $U/t=-10$ is shown in~\cref{fig:fflo}. For the non-polarized system, $\tilde{\psi}_0$ peaks for $k_1+k_2=0$, i.e., on the line of zero total momentum. However, if we introduce a spin-imbalance, the condensate wave function peaks on the off-diagonals $k_1+k_2=\pm q$, implying non-zero momentum in the $x$-direction. This indicates the formation of an FFLO state with finite total momentum, $q\neq 0$. Since we have established that the pairing is predominantly local in real space and of s-wave type, we can lower the computational cost by instead projecting on the local component of the $2$RDM  with $P_{\text{loc}}=\delta_{|\bm{r}_1 - \bm{r}_2|, 0}$, or explicitly,
\begin{align}
\label{eq:local2rdm}
    \rho_2^{\text{loc}}(\br,\br')&= \text{Tr}_{\br_2,\br_2^\prime}\left[P^\dagger \rho^S_2 P \right]
    \\
    &= \langle c^\dagger_{\br\uparrow} c^\dagger_{\br\downarrow} c_{\br'\downarrow} c_{\br'\uparrow} \rangle \\
    &= \sum_{l} \varepsilon_{l} \, \psi^\text{loc}_{l}(\br)^* \psi^\text{loc}_{l}(\br'),
\end{align}
which allows us to study larger systems. We show the spectra in~\cref{fig:fflo_momentum} (a). For visualization purposes, we study the rung averaged order parameter (or equivalently, projecting on $k_y=0$) of this component
\begin{equation}
    \bar{\psi}_0(x) = \frac{1}{W}\sum_y \psi^{\text{loc}}_0(x,y).
\end{equation}
This is shown in~\cref{fig:fflo_momentum} (c), together with its Fourier transform 
\begin{equation}
    \bar{\psi}_0(k) = \frac{1}{\sqrt{N}}\sum_x e^{-ikx}\bar{\psi}_0(x),
\end{equation}
see~\cref{fig:fflo_momentum} (b), for some values of magnetization $m$. We observe that the momentum increases as the magnetization is increased. It is also clear, since the order parameter is real, that this is a state of LO type, as previously reported in~\cite{luscher08}.

\subsection{\textit{d}-wave superconducting stripe order}

We now turn our attention to the repulsive Hubbard model, where yet other forms of superconductivity have been reported. At small doping, several recent studies have reported a coexistence regime between a superconductor and a charge density wave~\cite{wietek22,baldelli_fragmented_2025,jiang24, xu24, vanhala18, mai22}, also referred to as a supersolid~\cite{leggett70}. Interestingly, the condensate is shown to be fragmented in this case, where the number of dominant eigenvalues in the 2RDM matches the number of charge density wave peaks~\cite{wietek22,baldelli_fragmented_2025}. Here, we focus on a particular parameter set of the $t-t'-U$ Hubbard model where $U/t=10$ and $t'/t=-0.2$ with hole doping $d=1/16$ which is considered a relevant parameter set for cuprate superconductors in the underdoped regime~\cite{xu24,neto16}, on a $32\times 4$ lattice. Additionally, previous studies have found co-existence between superconductivity and stripe order, as well as condensate fragmentation~\cite{xu24,baldelli_fragmented_2025}. The ground state is computed at a bond dimension of $D=4000$. We assume nearest neighbour pairing and compute
\begin{align}
\label{eq:nn2rdm}
    \rho^S_2(\br\bm\mu: \br'\bm\mu') &= \Tr_{\br_2\br_2'}[P_{\mu}^\dagger \rho_2^S P_{\mu'}] \\ &= \braket{\Delta_{\br,\br+\bm\mu}\Delta_{\br',\br'+\bm\mu'}}
\end{align}
where $P_{\mu}=\delta_{|\br_1-\br_2|,\mu}$, with $\mu=|\bm\mu|$, and $\bm\mu,\bm\mu'=\hat{\bm x},\hat{\bm y}$ are the nearest neighbour vectors. The spectrum is shown in~\cref{fig:fragmented_momentum} (a). We observe the spectrum to be consistent with that of a fragmented condensate.

Since the pairing is known to be of $d_{x^2-y^2}$ type for these parameters~\cite{baldelli_fragmented_2025}, we compute 
\begin{equation}
    \bar{\psi}_l^{(d)}(x)=\frac{1}{4W}\sum_y \sum_{\bm{\mu}=\pm \hat{\bm{x}},\pm \hat{\bm{y}}}(-1)^{\bm{\mu} \cdot \hat{\bm{x}}}\psi_l((x,y),\bm{\mu})
\end{equation}
where $\bm{\mu}$ is a nearest-neighbour vector, and $\psi((x,y),\bm\mu)=\psi(\br,\br+\bm\mu)$. The condensate wave functions are shown in momentum and real space in~\cref{fig:fragmented_momentum} (b) and (c). In (c), we also show the rung-averaged hole density $n_h$, which forms a CDW. The fragmented condensates exhibit well defined momenta, with remarkable similarity to the FFLO states in the spin-imbalanced attractive Hubbard model. The wave function $\psi_{l=0}$ of the dominant eigenvalue peaks at $k=0$ and is largely uniform, while for greater $l$, the wave functions peak at $k\neq 0$, and has $\sum_x \bar{\psi}_{l\neq 0}^{(d)}(x)\approx 0$.

\section{Condensate composition}

\begin{figure}
    \centering
    \includegraphics[width=\linewidth]{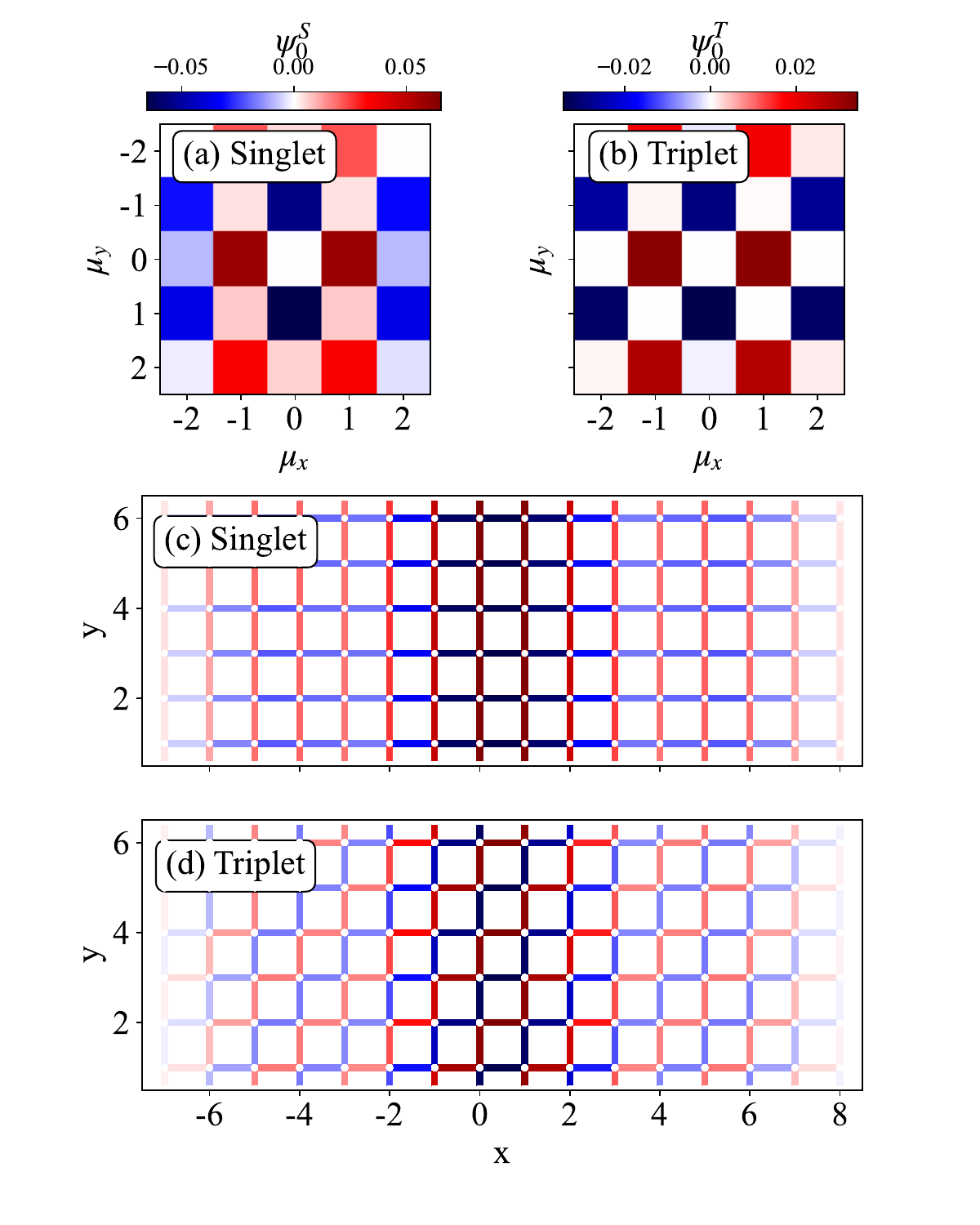}
    \caption{The singlet and triplet components of the dominant condensate wave function, $\psi_0$, for the repulsive $t-t'-U$ Hubbard model at $U/t=10$, $t'/t=0.2$, and $p=1/16$, on a $16\times6$ cylinder. (a) $\psi^S_0(\bm0,\bm\mu)$ for the singlet component with $\bm\mu$ the relative vector between the electrons. (b) $\psi^T_0(\bm0,\bm\mu)$ for the triplet component with $\bm\mu$ the relative vector between the electrons. (c) Nearest-neighbour part of the singlet component $\psi_0^S(\br,\br+\hat{\bm\mu})$, with $\hat{\bm\mu}=\hat{\bm x},\hat{\bm y}$ the nearest neighbour lattice vectors. (d) Nearest-neighbour part of the triplet component $\psi_0^T(\br,\br+\hat{\bm\mu})$, with $\hat{\bm\mu}=\hat{\bm x},\hat{\bm y}$ the nearest neighbour lattice vector.  As predicted by the projection, we see a $d$- and $p$-wave pattern in the singlet and triplet wave function respectively. We also visually observe the wave function for each relative vector $\bm\mu$, where in both cases the greatest support is on nearest-neighbour and $4$th nearest-neighbour pairing.}
    \label{fig:singlet_triplet}
\end{figure}

\begin{table}[]
    \centering
\begin{tabular}{|c|c||w{c}{0.9cm}|w{c}{0.9cm}|w{c}{0.9cm}|w{c}{0.9cm}|}
    \hline
    \multicolumn{1}{|c}{}  & \multicolumn{1}{c||}{} & \multicolumn{4}{c|}{$d$} \\
    \cline{3-6}
    \multicolumn{1}{|c}{}  & \multicolumn{1}{c||}{} & 0 & 1 & 2 & 3 \\
    \hhline{|=|=#=|=|=|=|}
    \multirow{5}{*}{Singlet} 
        & $\mathrm{A}_1$ & 0.949 & 0.045 & 0.001 & 0.001 \\
        \cline{2-6}
        & $\mathrm{A}_2$ & 0 & 0 & 0 & 0 \\
        \cline{2-6}
        & $\mathrm{B}_1$ & 0 & 0 & 0 & 0 \\
        \cline{2-6}
        & $\mathrm{B}_2$ & 0 & 0 & 0 & 0 \\
    \hline
\end{tabular}
    \caption{Attractive Hubbard model at $U/t=-10$ and $n=1/2$, on a $16\times4$ subsection of a $32\times4$ cylinder. The support of the condensate wave function $\psi_0$ on different point group irreps $\nu$, and distance index $d$. The attractive Hubbard model has the largest support on the $\mathrm{A}_1$ representation of the lattice point group, which is identifiable with $s$-wave pairing. Additionally, the pairing is shown to be local.}
    \label{tab:attractive}
\end{table}

\begin{table}[]
    \centering
\begin{tabular}{|c|c||w{c}{0.9cm}|w{c}{0.9cm}|w{c}{0.9cm}|w{c}{0.9cm}|w{c}{0.9cm}|}
    \hline
    \multicolumn{1}{|c}{}  & \multicolumn{1}{c||}{} & \multicolumn{5}{c|}{$d$} \\
    \cline{3-7}
    \multicolumn{1}{|c}{}  & \multicolumn{1}{c||}{} & 0 & 1 & 2 & 3 & 4 \\
    \hhline{|=|=#=|=|=|=|=|}
    \multirow{5}{*}{Singlet} 
        & $\mathrm{A}_1$ & 0 & 0 & 0.001 & 0 & 0.002 \\
        \cline{2-7}
        & $\mathrm{A}_2$ & 0 & 0 & 0 & 0 & 0 \\
        \cline{2-7}
        & $\mathrm{B}_1$ & 0 & 0.322 & 0 & 0.004 & 0.182 \\
        \cline{2-7}
        & $\mathrm{B}_2$ & 0 & 0 & 0 & 0 & 0 \\
    \hline
    \multirow{1}{*}{Triplet} 
        & $\mathrm{E}$   & 0 & 0.093 & 0 & 0 & 0.137 \\
    \hline
\end{tabular}
    \caption{Repulsive $t-t'-U$ Hubbard model at $U/t=10$, $t'/t=0.2$, and $p=1/16$, on a $16\times6$ cylinder. The support of the condensate wave function $\psi_0$ on different spin configurations, point group irreps $\alpha$, and distance index $d$. For the $t-t'-U$ Hubbard model, the singlet condensate wave function has the largest components on nearest-neighbour pairing in the $\mathrm{B}_1$ representation, and the triplet condensate wave function has all components in the $\mathrm{E}$ representation. We identify this with $d_{x^2-y^2}$ and $p$-wave pairing, respectively.}
    \label{tab:repulsive}
\end{table}

In the previous section, we assumed the pairing to be of singlet nearest-neighbour $d_{x^2-y^2}$-pairing type. In this section we justify this by studying the leading eigenvector of the full $2$RDM. We study the leading eigenvector of the full $2$RDM, $\rho_2$, for the repulsive $t-t'-U$ Hubbard model at $U/t=10$, $t'/t=0.2$, and $p=1/16$, on a $16\times6$ square lattice, with a bond dimension of $D=5000$. We also compare with the dominant eigenvector of the singlet $2$RDM, $\rho^S_2$, for the attractive Hubbard model at $U/t=-10$ and $n=1/2$, on a $16\times4$ subsection of a $32\times 4$ cylinder. 
First, we decompose $\psi_0$ into its spin quantum numbers, $s$ and $m$, using the projector onto singlets $P_S$, cf. \cref{eq:singletprojector} and on triplets
\begin{equation}\label{eq:tripletprojector}
    P_T = \frac{1}{4}(3I+\vec{\sigma}_1\cdot\vec{\sigma}_2).
\end{equation}
Doing so, we find that the leading eigenvector for the repulsive Hubbard model is mostly singlet, with a component of $0.663$, but also triplet ($m=0$), with a component of $0.337$. There are no components for the $m=\pm1$ triplets. This has previously been reported in the $t-t'-J$ model~\cite{jiang21}. Moreover, we project the wave function onto fixed pairing distances $\mu$, implemented by the projector \begin{equation}
    P_\mu = \delta_{|\br_1-\br_2|,\mu},
\end{equation}
as well as onto a particular irrep of the lattice point group, using~\cref{eq:projector_irrep} 
\begin{equation}
    P_\alpha \psi_0 = \frac{d_\alpha}{|G|}\sum_{g\in G} \chi_\alpha^*(g)\pi(g)\psi_0.
\end{equation}
Hence, in total we study
\begin{equation}
    |c_{\alpha,\mu,S/T}|^2=\lVert P_{\alpha}P_{\mu}P_{S/T}\psi_0\rVert^2,
\end{equation}
which is the component of the condensate wave function that transforms under the $\alpha$ irrep of the lattice point group, has a pairing distance $\mu=|\br_1-\br_2|$ between the two electrons, and is either a singlet or a triplet. Since the set $\{P_{\alpha}P_{\mu}P_{S/T}\}$ is a complete set of orthogonal projectors, we have,
\begin{equation}
    \sum_{\nu,\mu,s=S/T} |c_{\alpha,\mu,s}|^2 = 1.    
\end{equation}
The component of each irrep $\alpha$ on $\psi_0$ is therefore $||P_{\alpha}\psi_0||^2$. Our lattice has a D$_4$ point symmetry, with the irreps denoted, using Mulliken notation~\cite{mulliken55}, as $\mathrm{A}_1$, $\mathrm{A}_2$, $\mathrm{B}_1$, $\mathrm{B}_2$, and $\mathrm{E}$. The action of $\pi(g)$ on $\psi_0$ is defined by letting the group element $\pi(g)$, i.e., rotations and reflections, act on the relative vector $\bm \mu=\br_2-\br_1$, with the wave function, $\psi_0(\br_1,\bm\mu)$, now being a function of one of the lattice sites, and the relative vector. This ensures the electrons in the pair remain on the lattice after e.g., a rotation or reflection. This works well for the case of the spin singlet. However, no such analysis is required for the triplet component $\psi_0^T$. Since $\psi_0^T$ is anti-symmetric under $\br_1 \leftrightarrow \br_2$, $\psi_0^T(\br_1\br_2)=-\psi_0^T(\br_2\br_1)$, which is equivalent to a rotation by $\pi$ around the center of mass, it transforms purely under the $\mathrm{E}$ representation, often referred to as $p$-wave pairing,
\begin{equation}
    P_E\psi_0^T(\br_1\br_2) = \frac{1}{2}\Big(\psi_0^T(\br_1\br_2) - \psi_0^T(\br_2\br_1)\Big) = \psi_0^T(\br_1\br_2).
\end{equation}
We also confirm this numerically. Conversely, the singlet component, $\psi_0^S$, is symmetric under $\br_1 \leftrightarrow \br_2$, and hence does not have a component in the $\mathrm{E}$ representation.

For the case of the attractive Hubbard model, see~\cref{tab:attractive}, we restrict ourselves to the spin-singlet, and recover the well-known results that the order parameter transforms in the trivial representation $\mathrm{A}_1$, identified with s-wave pairing. This can be compared with the case of superconductivity in the repulsive $t-t'-U$ Hubbard model, see~\cref{tab:repulsive}. The leading eigenvector transforms primarily in the $\mathrm{B}_1$ representation of D$_4$, identified with $d_{x^2-y^2}$-wave pairing. Importantly, we also find the strongest support on nearest-neighbour spin-singlet pairing, justifying our previous computations. 
Motivated by these results, we show the nearest neighbour condensate wave functions for the singlet and triplet in~\cref{fig:singlet_triplet} (a,c), as well as the dependence of the pairing of the relative vector in~\cref{fig:singlet_triplet} (b,d). For the singlet, we observe the expected $d$-wave pattern, and for the triplet we observe the $p$-wave "staircase"-pattern. Thus we visually confirm the results given in~\cref{tab:repulsive}.

\section{Scaling properties of the $2$RDM} \label{sec:circulant}

Let us consider a marginalized and projected 2RDM of the form $\rho_2(\br, \br^\prime)$, for example the local 2RDM, $\rho^{\text{loc}}_2$, as defined in \cref{eq:local2rdm} or $\rho^S_2$ as in \cref{eq:nn2rdm}. With translation invariance, this correlation matrix only depends on the relative distances,
\begin{equation}
    \rho_2(\br,\br')=C(\br-\br').
\end{equation}
In one dimension, the correlation matrix $\rho_2(\br,\br')$ is a Toeplitz matrix. With periodic boundary conditions on a finite lattice of length $L$, it is also a circulant matrix~\cite{gray06,davis79}, i.e., of the form $C_{ij}=c_{(i-j)\mod{L}}$ where $L$ is the system size,
\begin{equation}
    C = \begin{pmatrix}
        c_0 & c_1 & \dots & c_{L-2} & c_{L-1} \\
        c_{L-1} & c_0 & c_{1} & \dots & c_{L-2} \\
        \vdots & c_{L-1} & c_0 & \ddots & \vdots \\
        c_{2} & \vdots & \ddots & \ddots & c_{1} \\
        c_{1} & c_{2} & \dots & c_{L-1} & c_0
    \end{pmatrix}.
\end{equation}
Circulant matrices are diagonalized by Fourier transformation with eigenvalues,
\begin{equation}
    \label{eq:circ_ev}
    \varepsilon_k = \sum_{j=0}^{L-1}c_j e^{-ijk}
\end{equation}
and eigenvectors, 
\begin{equation}
    \label{eq:fourier_vec}
    \bm\psi^\text{F}_k=\frac{1}{\sqrt{L}}(1,\omega_k,\omega^{2}_k,\dots,\omega^{(L-1)}_k)^T\, ,
\end{equation}
where $\omega_k=e^{ik}$
and $k=0,2\pi/L,\dots , 2\pi(L-1)/L$. The eigenvalues are real whenever $C$ is Hermitian, i.e. $C(\br)^* = C(-\br)$.

Let us assume algebraically decaying correlation functions with periodic boundary conditions of the form, 
\begin{equation}
    c_j=\frac{1}{2}\left(j^{-K_{sc}} + (L-j)^{-K_{sc}} \right) e^{iq j },
\end{equation}
where $K_{sc}$ denotes the exponent of algebraic decay and we also take into account a possible periodic modulation with wave number $q$. This scenario is relevant for Luther-Emery liquids~\cite{luther-emery}, often encountered on cylinder geometries in DMRG. The terms in the sum, \cref{eq:circ_ev}, interfere constructively only for $k=q$. Hence, there is only one dominant eigenvalue at wave number $q$,
\begin{equation}
\begin{split}
    \varepsilon_q &= \sum_{j=0}^{L-1}\frac{1}{2}\left(j^{-K_{sc}} + (L-j)^{-K_{sc}} \right) \\
    &\approx \int_0^{L}dx \; x^{-K_{sc}}
    \propto L^{1-K_{sc}},
\end{split}
\end{equation}
which shows that the dominant eigenvalue grows like $L^{\nu}$ with $\nu=1-K_{sc}$. In the case of genuine ODLRO, $K_{sc}=0$, and thus $\nu=1$. Since the number of electrons, $N_e$, is proportional to the system size $L$ for fixed density, this scaling behaviour is precisely the Penrose-Onsager criterion for the formation of a condensate, i.e. $\varepsilon_q \sim N_e$. 

This result is readily generalized to higher dimensions. In two dimensions, we can write the compound index $i=(i_x-1)L_y+i_y$, and the resulting matrix $C$ is a block-circulant matrix~\cite{tee05}, i.e., it takes the form
\begin{equation}
    C = \begin{pmatrix}
        B_0 & B_1 & \dots & B_{L_x-2} & B_{L_x-1} \\
        B_{L_x-1} & B_0 & B_{1} & \dots & B_{L_x-2} \\
        \vdots & B_{L_x-1} & B_0 & \ddots & \vdots \\
        B_{2} & \vdots & \ddots & \ddots & B_{1} \\
        B_{1} & B_{2} & \dots & B_{L_x-1} & B_0
    \end{pmatrix},
\end{equation}
where $B_i$ are $L_y\times L_y$ matrices. Additionally, if we have periodic boundary conditions in both directions, the blocks $B_i$ themselves are circulant matrices. The matrix is thus diagonalized by a two dimensional Fourier transform, and a similar analysis as above can be made. 

\section{Supersolids}\label{sec:supersolids}

\begin{figure}
    \centering
    \includegraphics[width=\linewidth]{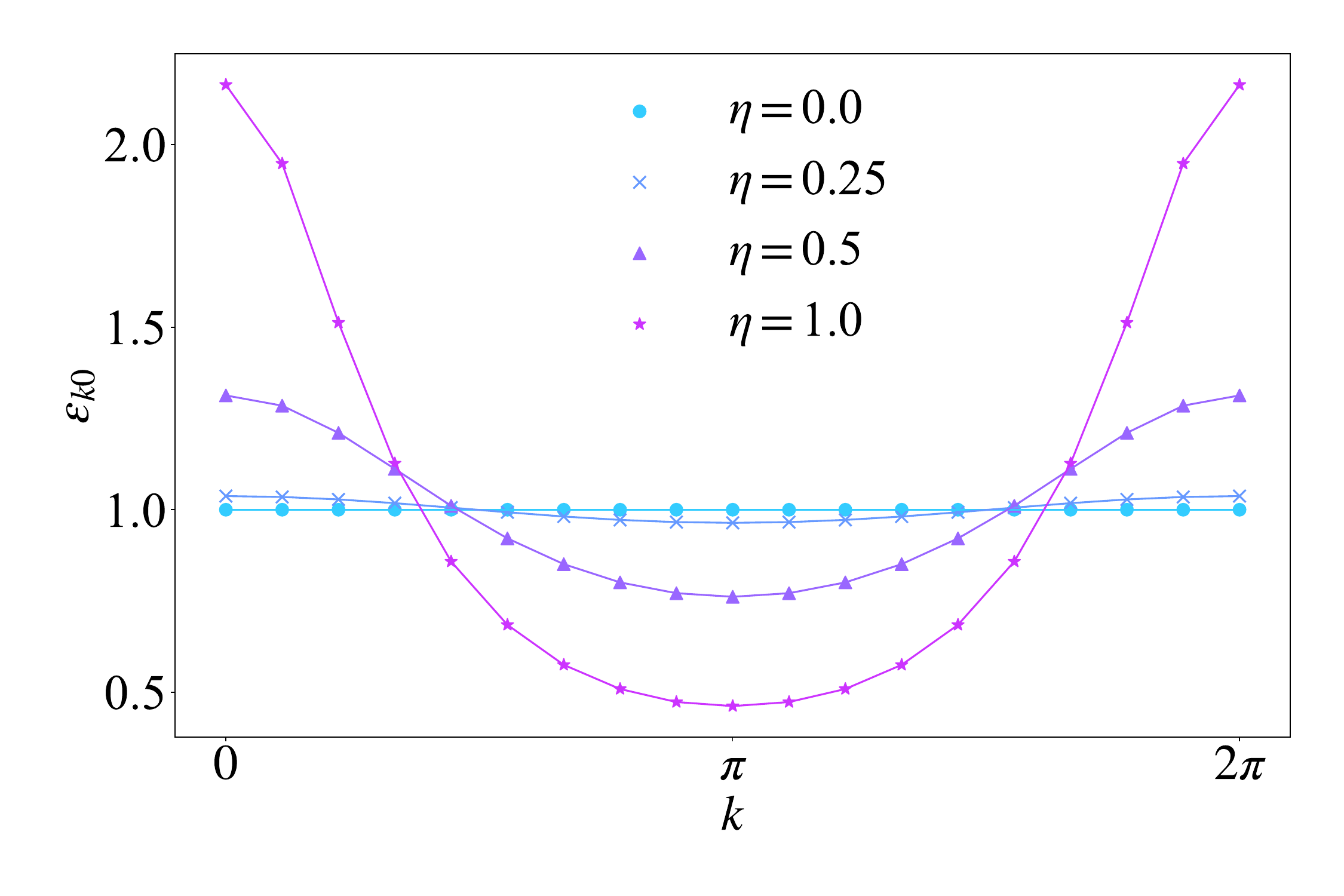}
    \caption{Eigenvalues of the block circulant matrix of form \cref{eq:2rdm_simplified}, with $g(s)=e^{-s/\eta}$, for different values of hopping parameter $\eta$. $\eta=0$ corresponds to decoupled CDW peaks. The eigenvalues as a function of momentum form a band, with the bandwidth $w$ increasing with $\eta$.}
    \label{fig:block_circulant}
\end{figure}

Now, let us consider a state where the translational symmetry is broken, by e.g., a CDW with period $M$, from $\mathbb{Z}_L$ down to $\mathbb{Z}_{S}$, where $S=L/M$. We employ a compound index $i=(i_s-1)M+i_m$, where $i_s=1,\dots,S$ and $i_m=1,\dots, M$ label the CDW peaks and internal sites, respectively. Since we have a symmetry of translations by $M$ sites, we can naturally model the correlation function as a block-circulant matrix $C_{ij}=(B_{i_s-j_s})_{i_mj_m}$, where the blocks $B$ only depend on $i_m$, $j_m$, and $i_s-j_s$. Similar to conventional circulant matrices, the eigenvalue computation of a block circulant matrix can be simplified using a Fourier transformation. Denoting $s=i_s-j_s$, the set of eigenvalues of a block-circulant matrix is the union of the eigenvalues of the matrices~\cite{tee05}
\begin{equation}\label{eq:eigenvalues}
    \Lambda_k = \sum_{s=0}^{S-1} B_s e^{iks},
\end{equation}
with $k=0,2\pi/S,\dots,2\pi(S-1)/S$, and the eigenvectors are 
\begin{equation}
    \bm{\Psi}_{kl} = \bm{\psi}^\text{F}_k \otimes \bm{\psi}^Q_{kl},
\end{equation}
where $\bm{\psi}^\text{F}_k$ is the Fourier vector, as in \cref{eq:fourier_vec}, and $\bm{\psi}^Q_{kl}$ are the eigenvectors of $\Lambda^k$.

Now, consider the case of a superconducting condensate on each CDW peak, with no correlation between the peaks. We should find $C$ to be block-diagonal, $C=\text{diag}(B,\dots,B)$, where $B$ is the $2$RDM describing the local superconducting condensate at each peak. This is also a block-circulant matrix where the diagonal entries are $B_0=B$, and the off-diagonal entries are zero, i.e. $B_s=0$ for $s\neq 0$. If $B$ is describing a simple condensate, this trivially yields one (degenerate) dominant eigenvalue, $\varepsilon_{0}$, per CDW peak. The eigenvectors will be localized in real space at the peaks. If we now allow the Cooper pairs to tunnel to the other wave peaks, we can approximate the on-peak contribution, $B_0=B$ as unchanged up to leading order, while $B_s$ for $s\neq 0$, become finite. If we only allow weak tunnelling, the contributions from $s\neq 0$ can be considered to be small perturbations. Using Equation \eqref{eq:eigenvalues}, we write
\begin{equation}
    \Lambda_k = B + \delta \cdot Q_k
\end{equation}
where $Q_k$ is a momentum dependent perturbation matrix, and $\delta$ is a scalar controlling the perturbation
\begin{equation}
    \delta \cdot Q_k = \sum_{s=1}^{S-1} B_s e^{iks} 
\end{equation}
Using matrix perturbation theory, the eigenvalues are then
\begin{equation}
    \varepsilon_{kl} =\varepsilon_{l}(B) + \delta \cdot \bm{\psi}^{s*}_lQ_k \bm{\psi}^s_l + \mathcal{O}(\delta^2),
\end{equation}
where $\varepsilon_{l}(B)$ and $\bm{\psi}^s_l$ are the $l$-th eigenvalue and eigenvector of $B$, respectively, with $l=0,\dots,M-1$. We can thus conclude that the eigenvalue degeneracy is lifted by the perturbation, which is allowed by Cooper pairs tunnelling between the CDW peaks. Moreover, we get one dominant eigenvalue $\varepsilon_{k0}$ per peak, labelled by momentum $k$. This is precisely what was observed for the fragmented condensates in the $t-t^\prime-U$ Hubbard model~\cite{baldelli_fragmented_2025}. Since $\varepsilon_0(B)$ grows linearly in the number of electrons per CDW peak, then so will also $\varepsilon_{kl}$. However, due to the CDW, if we let the system grow in the $y$-direction, each eigenvalue will grow linearly in size, while if the system grows in $x$, we will instead introduce more leading eigenvalues and condensates.

Moreover, the total condensate fraction remains conserved up to leading order, because
\begin{multline}
    \sum_k \varepsilon_{k0} = \sum_k (\varepsilon_{0}(B) + \delta \cdot \bm{\psi}^{s*}_0Q_k \bm{\psi}^s_0) = \\ S\varepsilon_{0}(B) + \delta \cdot \bm{\psi}_0^{s}\left(\sum_k Q_k \right) \bm{\psi}_0 = S\varepsilon_{0}(B),
\end{multline}
which is independent of the perturbation $\delta \cdot Q_k$, since $\sum_k Q_k=0$.

As an example, let us assume
\begin{equation}\label{eq:g(s)}
    B_s = g(s)B,
\end{equation}
where $g(s)$ is a decaying function in $s$. Equation \eqref{eq:eigenvalues} then simplifies to 

\begin{equation}\label{eq:2rdm_simplified}
    \Lambda_k = B\sum_{s=0}^{S-1}g(s)e^{iks}.
\end{equation}
If we let $\varepsilon_0$ be the macroscopic eigenvalue of $B$, then the macroscopic eigenvalues of $\Lambda_k$ form a band $\varepsilon_{k0}$. For 
\begin{equation}\label{ansatz}
    g(s)=e^{-s/\eta}
\end{equation} 
with $\eta$ a characteristic length, the macroscopic eigenvalues $\varepsilon_{k0}$ of $\rho_2$ are shown in~\cref{fig:block_circulant} (normalized to $\varepsilon_{0}=1$). It is clear that for $\eta=0$ i.e., no interaction between the stripes, that the macroscopic eigenvalue is degenerate for all the CDW peaks. When we increase the value of $\eta$ though, the degeneracy is lifted, and we see a band of eigenvalues form where each eigenvalue has a well-defined momentum. We also see how the bandwidth, $w=\varepsilon_{00}-\varepsilon_{\pi0}$, grows as we increase $\eta$. In the other limit, $\eta\to \infty$, we end up with $\Lambda_k=BS\delta_{k,0}$, and recover a simple condensate at momentum $k=0$. Moreover, the bandwidth increases with increasing tunnelling between the stripes. As such, a large spread of dominant eigenvalues in a stripe-fragmented superconductor is an indication of enhanced coherence across the stripes.

Concerning the bandwidth, the difference between the largest and smallest dominant eigenvalue, it is found to be
\begin{equation}
    w = \varepsilon_{00}-\varepsilon_{\pi0} = 2\psi_0^{s*}\Big(\sum_{s \text{ odd}} B_s\Big)\psi_0^s,
\end{equation}
assuming all $B_s$ are positive semi-definite. 

If we assume the form of~\cref{eq:g(s)}, relevant for a CDW in the $x$-direction, then 
\begin{equation}
    w=2\varepsilon_0(B)\left( \sum_{s\text{ odd}} g(s) \right),
\end{equation}
scales linearly with $L_y$, while the scaling with $L_x$ depends on the form of $g(s)$. For example, if $g(s)$ decays exponentially (\cref{eq:g(s)}), $w$ is independent of $L_x$ for large $L_x\gg \eta$.

\section{Discussion}
We proposed a versatile and accurate formalism for the investigation of superconducting condensates based on the study of the 2RDM $\rho_2$ as defined in \cref{eq:full2rdm}. The formation of a simple condensate is diagnosed by a linear scaling of the leading eigenvalue, encoding the condensate fraction, with the number of particles in the system~\cite{Leggett,Leggett2022}. This fact is also known as the Penrose-Onsager criterion for condensate formation~\cite{penrose-onsager,baldelli_fragmented_2025}. Moreover, the approach also allows for the detection of fragmented condensates where two or more leading eigenvalues scale with the number of particles, which cannot be achieved by diagnosing the presence of off-diagonal long-range order alone~\cite{Leggett}. 

We then gave a precise definition of the symmetries of the condensate and showed that a linear scaling in the number of particles of a projected 2RDM implies a linear scaling in the full 2RDM, proved using the Poincar\'e separation theorem. Applied to projectors onto irreps of the symmetry group, we defined $\alpha$-condensates ($\alpha$ denotes an irrep) to be condensates where the leading eigenvalue of the $\alpha$-projected 2RDM scales linearly in the number of particles. 

The predictive power of the approach was then showcased by applications to several well-studied cases of superconducting states in the two-dimensional Hubbard model. The attractive Hubbard model on a square lattice realizes a superconducting ground state \cite{scalettar89,moreo91,singer98}. Since attractive interactions allow for sign-problem-free auxiliary-field Quantum Monte Carlo simulations, we were able to demonstrate a clean linear scaling of the condensate fraction with system sizes up to $20 \times 20$, serving as highly conclusive evidence for the formation of a condensate at quarter-filling for $U/t=-4$ and $U/t=-10$. By further investigation of the corresponding eigenvector, i.e., the Cooper pair wave function, we demonstrated that the pairing is predominately of on-site $s$-wave character and localized near the non-interacting Fermi surface in momentum space. Although qualitatively these results were previously known, our approach allowed for an unprecedented accuracy in measuring the precise properties of the Cooper pairs. In particular, we investigated the localization properties by studying the wave function variance and inverse participation ratios across the BCS-BEC crossover upon scanning the value of $U/t$. Extrapolated results to the thermodynamic limit are compared to those of BCS mean-field theory, where significant quantitative differences were found. 

As not every superconducting phase of the 2D Hubbard model can be efficiently simulated using QMC due to the sign-problem, we compared the 2D limit to simulations performed using DMRG on cylinders. On quasi-one-dimensional cylinders, the scaling of the leading eigenvalue is found to scale algebraically with an exponent $\nu < 1$, which is directly related to the conventionally studied superconducting Luttinger parameter $K_{sc}=1-\nu$. In a finite (Zeeman) magnetic field, the attractive Hubbard model enters an FFLO phase, for which a sign-problem-free QMC formulation is unknown. Applying our analysis to this state using DMRG cylinders, we were able to track the finite momentum of the Cooper pairs accurately, and compare the condensate fractions to the uniform $s$-wave superconductor in zero magnetic field. 

Next, we turned our attention to the supersolid states in the repulsive Hubbard model, where superconductivity and CDWs coexist. As previously reported~\cite{wietek22,baldelli_fragmented_2025}, we find a fragmentation of the condensate, where the fragments are Bloch waves in the super-unit cell of the CDW, and there is exactly one condensate fragment per period of the CDW. However, by carefully studying the full 2RDM, also the triplet component, we revealed that a significant contribution ($\sim 30 \%$) to the condensate fraction comes from a triplet $p$-wave superconducting state. This remarkable finding now questions our understanding of the superconducting state encountered in the repulsive Hubbard model, which was previously assumed to be purely singlet and $d$-wave, except for one previous study pointing out this coexistence in the $t$-$t^\prime$-$J$ model~\cite{jiang21}.
This leaves interesting questions for future studies: (i) How does the triplet component of the dominant condensate scale with system size? (ii) Will there still be a finite contribution of the triplet sector in the full two-dimensional limit, either as a coherent mixture of a single condensate or as two separate states of the fragmented condensate? (iii) It would be highly interesting to understand the interplay between a condensate consisting of both singlet and triplet Cooper pairs with an antiferromagnetic spin order; i.e., can the spin correlations induce the mixture of singlet and triplet components?

Finally, we discussed how a periodic modulation of pairing correlations naturally can lead to condensate fragmentation, and showed analytically that then the fragments indeed carry momentum quantum numbers in units of the super-unit cell of the modulation wavelength. This analysis suggests a simple analogy: The Cooper pairs in the supersolid state can be thought of as particles hopping on the lattice defined by the CDW. We showed, that increasing the tunneling between neighboring modulations leads to an increase in the bandwidth of the fragments, which can be directly probed by studying the spread of the leading eigenvalues of the 2RDM. Thus, we understood that the dominant eigenvalues in a supersolid with weakly coupled CDW periods (i.e., weak pairing correlations) will be only weakly split, whereas strong coupling leads to a larger splitting between the leading eigenvalues.

Our approach also removes a specific bias from numerical studies of superconducting states: the bias of choosing measurements. If one directly measures the full 2RDM as in \cref{eq:full2rdm}, the components of the condensate can be mechanically extracted by using proper projectors onto all irreps of the symmetry group. As demonstrated for the supersolid state in the repulsive Hubbard model, this approach not only revealed the fragmented nature of the condensate, but also showed a significant triplet $p$-wave contribution not discussed widely before.

Here, we have used our framework in conjunction with two numerical techniques, AFQMC and DMRG. However, since our approach is generic, it can easily be adopted by other numerical and analytical techniques, such as variational Monte Carlo or recent neural quantum state methods and argue that it allows for an accurate and unbiased characterization of superconducting states of matter in strongly correlated electron systems.

\begin{acknowledgments}

We thank Aritra Sinha, Chris Hooley, Parasar Thulasiram, and Dhruv Tiwari for useful discussions.
A.W. acknowledges support by the German Research Foundation (DFG) through the Emmy Noether program (Grant No. 509755282) and the European
Research Council (ERC) under the European Union’s Horizon
Europe research and innovation program (Project ID 101220368)—ERC Starting Grant MoNiKa.
J.H. acknowledges financial support by the Deutsche Forschungsgemeinschaft (DFG, German Research Foundation) through the Würzburg-Dresden Cluster of Excellence ctd.qmat – Complexity, Topology and Dynamics in Quantum Matter (EXC 2147, project-id 390858490)
and via the project A07 of the Collaborative Research Center SFB 1143 (Project No. 247310070).
The auxillary field QMC simulations were carried out with the ALF package~\cite{ALF} available at \url{https://alf.physik.uni-wuerzburg.de}. The DMRG simulations were obtained using the ITensor Library \cite{itensor,itensor-r0.3}.

\end{acknowledgments}

\appendix

\section{Details on AFQMC simulation} \label{app:afqmc}

\begin{figure}[t]
    \centering
    \includegraphics[width=\linewidth]{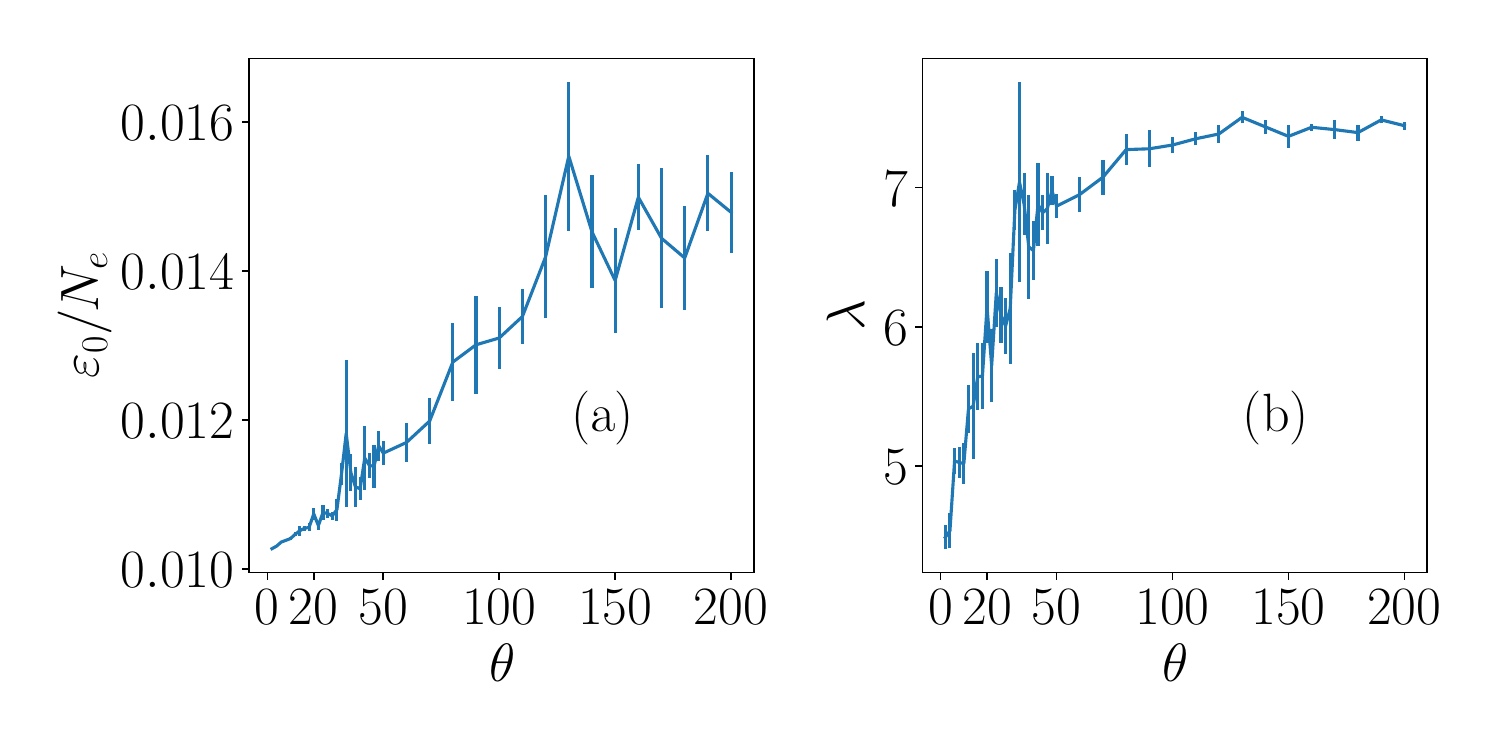}
    \caption{Convergence plots in projection time $\theta$ for AFQMC simulations on the attractive Hubbard model with $U/t=-1$ and $n=0.5$ on a $20\times20$ square lattice. (a) The dominant eigenvalue $\varepsilon_0/N_e$ and (b) the localization length as a function of projection time $\theta$.}
    \label{fig:U(-1.0)_L(20)_n(0.5)}
\end{figure}

\begin{figure}[t]
    \centering
    \includegraphics[width=\linewidth]{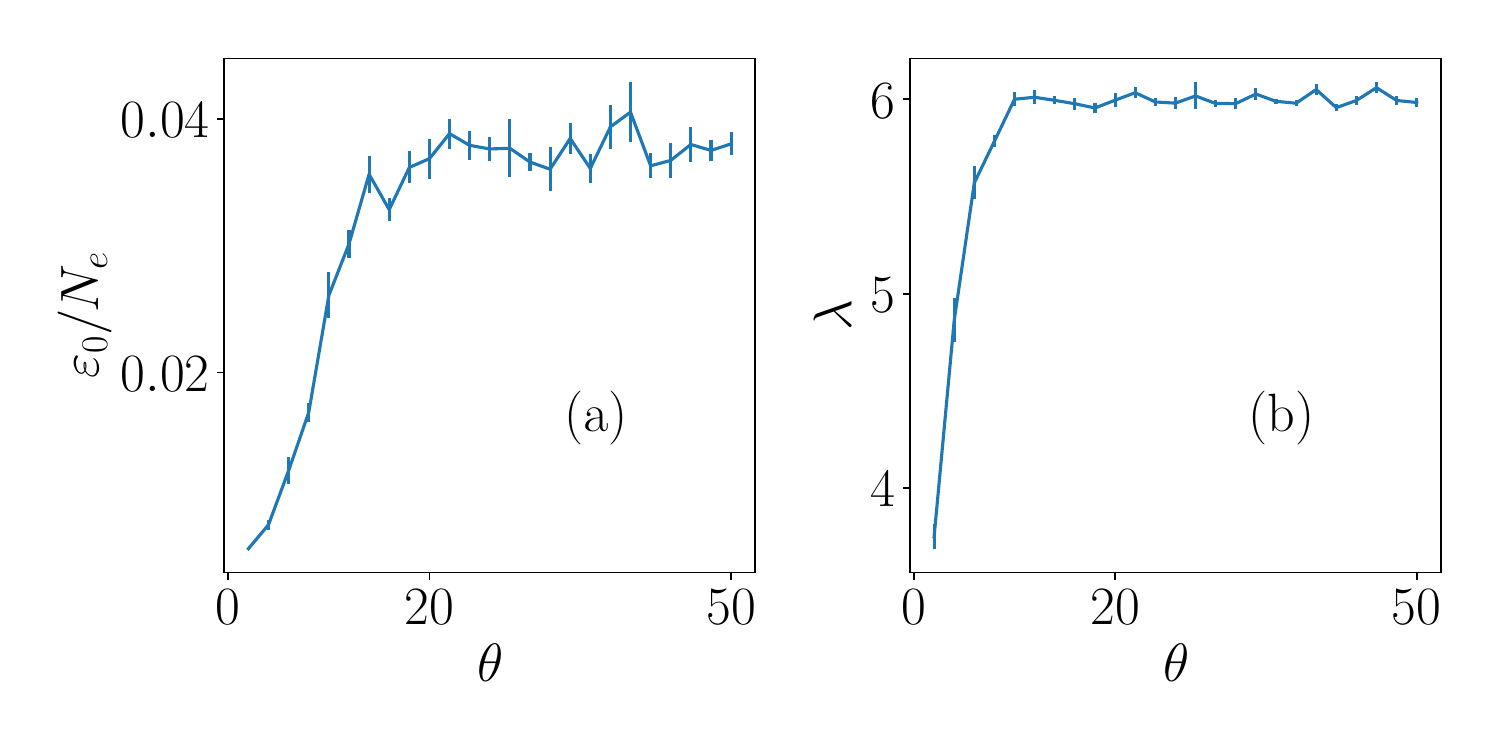}
    \caption{Convergence plots in projection time $\theta$ for AFQMC simulations on the attractive Hubbard model with $U/t=-1$ and $n=1.0$ on a $20\times20$ square lattice. (a) The dominant eigenvalue $\varepsilon_0/N_e$ and (b) the localization length as a function of projection time $\theta$.}
    \label{fig:U(-1.0)_L(20)_n(1.0)}
\end{figure}

\begin{figure}[t]
    \centering
    \includegraphics[width=\linewidth]{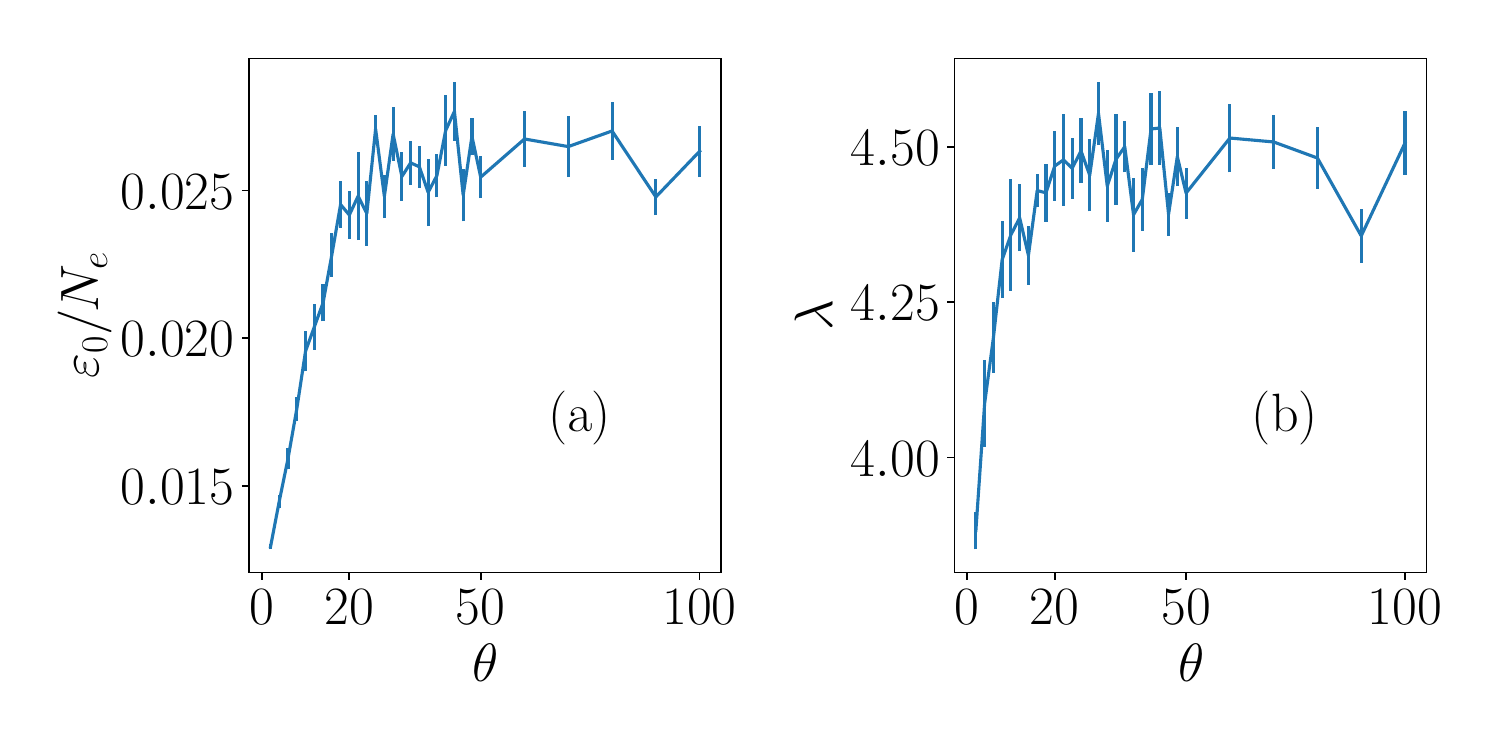}
    \caption{Convergence plots in projection time $\theta$ for AFQMC simulations on the attractive Hubbard model with $U/t=-2$ and $n=0.5$ on a $20\times20$ square lattice. (a) The dominant eigenvalue $\varepsilon_0/N_e$ and (b) the localization length as a function of projection time $\theta$.}
    \label{fig:U(-2.0)_L(20)_n(0.5)}
\end{figure}

We employ the usual auxiliary-field Quantum Monte Carlo algorithm as it is implemented in the `Algorithm for Lattice Fermions' (ALF) software package \cite{ALF}, supplemented by the new observable $\rho_2^{\bm{q}=0,S}$. The interaction is decomposed in the $S^z$ channel via Hubbard-Stratonovich transformations; the Trotter step size is set to $\Delta\tau=0.1$. The trial wave function is determined from a non-interacting Hamiltonian, $H_{\mathrm{trial}}$, by occupying the lowest $N_e/2$ states in the spin-down sector while the spin-up part is the time-reversal conjugate. The non-interacting Hamiltonian $H_{\mathrm{trial}}$ is given by electrons on a square lattice with dimerized nearest-neighbour hopping (see Sec.~7.6 of Ref.~\cite{ALF}. At half filling, the dimerisation opens a small gap on the Fermi surface such that the trial wave function is given by the unique ground state. At quarter filling, the corresponding eigenstates of $H_{\mathrm{trial}}$ are degenerate such that the trial wave function slightly breaks translation symmetry while maintaining time-reversal symmetry. We note that the superconducting ground states preserve both translation and four-fold rotation symmetry and the `broken' symmetries of the trial wave functions are restored upon projection. In particular, the trial wave function therefore cannot act as a `pinning field'.

Convergence plots of the dominant eigenvalue $\varepsilon_0$ and the localization length $\lambda$ are shown for a set of parameters in~\cref{fig:U(-1.0)_L(20)_n(0.5),fig:U(-1.0)_L(20)_n(1.0),fig:U(-2.0)_L(20)_n(0.5)}. In~\cref{fig:U(-1.0)_L(20)_n(0.5)}, we see $U/t=-1$, $n=0.5$, and $L=20$. Out of the studied parameters, this was the most challenging, converging first around $\theta=150$. In~\cref{fig:U(-1.0)_L(20)_n(1.0)}, we see $U/t=-1$, $n=1$, and $L=20$, which converged significantly quicker, around $\theta=20$. Finally, in~\cref{fig:U(-2.0)_L(20)_n(0.5)}, we show $U/t=-2$, $n=0.5$, and $L=20$. Here we find the observables to be well converged at $\theta=50$. For all other sets of parameters, we found $\theta=20$ to be enough for convergence.

\section{Details on mean-field calculations} \label{app:mf}

 Within the Cooper channel, the mean-field Hamiltonian of the Hubbard model in momentum space is
\begin{equation}
    H = \sum_{\bk\sigma} \varepsilon_{\bk} c_{\bk\sigma}^\dagger c_{\bk\sigma} + \sum_{\bk} \Delta c_{\bk\uparrow}^\dagger c_{-\bk\downarrow}^\dagger + \Delta^* c_{-\bk\downarrow} c_{\bk\uparrow}
\end{equation}
with $\varepsilon=-t\cos{\bk_x} - t\cos{\bk_y}$, $\Delta=\frac{U}{V}\sum_{\bk}\langle c_{-\bk\downarrow} c_{k\uparrow} \rangle$, and $V$ the volume of the system. Using Wick's theorem, we can evaluate the $2$RDM as
\begin{equation}
    \begin{split}
    \rho_2(\bk_1,\bk_2,\bk_3,\bk_4) = \langle c_{\bk_1 \sigma_1}^\dagger c_{\bk_2 \sigma_2}^\dagger c_{\bk_4 \sigma_4} c_{\bk_3 \sigma_3} \rangle = \\
     \langle c_{\bk_1\sigma_1}^\dagger c_{\bk_3\sigma_3} \rangle \langle c_{\bk_2\sigma_2}^\dagger c_{\bk_4\sigma_4} \rangle \\
    - \langle c_{\bk_1\sigma_1}^\dagger c_{\bk_4\sigma_4} \rangle \langle c_{\bk_2\sigma_2}^\dagger c_{\bk_3\sigma_3} \rangle \\
    + \langle c_{\bk_1\sigma_1}^\dagger c_{\bk_2\sigma_2}^\dagger \rangle\langle c_{\bk_4\sigma_4}^\dagger c_{\bk_3\sigma_3}^\dagger \rangle
    \end{split}
\end{equation}
since the mean-field theory is Gaussian. We can now evaluate the individual terms using results from BCS theory, which has the variational state
\begin{equation}
    \ket{\text{BCS}} = \prod_{\bm{q}} (u_{\bm{q}}+v_{\bm{q}}c^\dagger_{\bm{q}\uparrow}c^\dagger_{-\bm{q}\downarrow})\ket{0},    
\end{equation}
which yields the singlet $2$RDM
\begin{equation} \label{eq:gaussian_2rdm}
\begin{split}
    \rho^{\bm q=0,S}_2(\bk,\bk') &= \frac{1}{2}\big( \delta_{\bk,\bk'}n_{\bk\uparrow}n_{-\bk\downarrow} + F_{\bk}^*F_{\bk'} \\
    &+ \delta_{\bk,-\bk'}n_{k\uparrow}n_{-\bk\downarrow} + F_{\bk}^*F_{-\bk'} \\
    &+ \delta_{\bk\bk'}n_{-\bk\uparrow}n_{\bk\downarrow} + F_{-\bk}^*F_{\bk'} \\
    &+ \delta_{\bk\bk'}n_{-\bk\uparrow}n_{\bk\downarrow} + F_{-\bk}^*F_{-\bk'}\big),
    \end{split}
\end{equation}
where 
\begin{equation}
    n_{\bk\uparrow} = n_{\bk\downarrow} = \frac{1}{2}\left( 1-\frac{\varepsilon_{\bk}-\varepsilon_F}{\sqrt{(\varepsilon_{\bk}-\varepsilon_F)^2 + \Delta^2}} \right),
\end{equation}
and
\begin{equation}
    F_{\bk} = \langle c_{-\bk\downarrow}c_{\bk\uparrow} \rangle =\frac{\Delta}{2\sqrt{(\varepsilon_{\bk}-\varepsilon_F)^2+\Delta^2}}.
\end{equation}
The gap $\Delta$ is found by self-consistently solving 
\begin{equation}
    \Delta = \frac{U}{V} \sum_{\bk} \frac{\Delta}{2\sqrt{(\varepsilon_{\bk}-\varepsilon_F)^2+\Delta^2}}.
\end{equation}
The $2$RDM is then constructed using Equation \eqref{eq:gaussian_2rdm} and diagonalized, yielding the dominant eigenvalue and condensate wave function.

\bibliography{main}

\end{document}